\documentclass[fleqn,usenatbib]{mnras}

\usepackage{newtxtext,newtxmath,xcolor}
\DeclareRobustCommand{\VAN}[3]{#2}
\let\VANthebibliography\thebibliography
\def\thebibliography{\DeclareRobustCommand{\VAN}[3]{##3}\VANthebibliography}
\newcommand{\bdb}{{\bf BdB}}
\newcommand{\ful}{$\mathcal{F}_{full}$}
\newcommand{\nomi}{$\mathcal{F}_{nom}$}
\newcommand{\radmm}{rad~m$^{-2}$}
\definecolor{DarkRed}{rgb}{0.25, 0.02, 0.02}
\newcommand{\rev}[1]{#1}
\newcommand{\revv}[1]{#1}
\usepackage{graphicx}	% Including figure files
\usepackage{amsmath}	% Advanced maths commands
\title[Full Resolution Faraday Spectra]{Full Resolution Deconvolution of Complex Faraday Spectra}

\author[L. Rudnick \& W. D. Cotton]{
Lawrence Rudnick,$^{1}$\thanks{E-mail: larry@umn.edu (LR)}
W. D. Cotton,$^{2}$\\
$^{1}$Minnesota Institute for Astrophysics, University of Minnesota, 116 Church St. SE, Minneapolis, MN 55455, USA\\
$^{2}$National Radio Astronomy Observatory, 520 Edgement Rd., Charlottesville, VA 22901, USA
}

\date{Accepted to MNRAS, 4 April, 2023.}
\pubyear{2023}

\begin{document}
\label{firstpage}
\pagerange{\pageref{firstpage}--\pageref{lastpage}}
\maketitle

\begin{abstract}

\rev{Polarized synchrotron emission from multiple Faraday depths can be separated by calculating the complex Fourier transform of the Stokes' parameters as a function of the wavelength squared, known as Faraday Synthesis. As commonly implemented, the transform introduces an additional term $\lambda_0^2$, which broadens the real and imaginary spectra, but not the amplitude spectrum.  We use idealized tests to investigate whether additional information can be recovered with a clean process restoring beam set to the narrower width of the peak in the real ``full" resolution spectrum with $\lambda_0^2=0$. We find that the $\lambda_0^2$ choice makes no difference, \emph{except for the use of a smaller restoring beam}. With this smaller beam, the accuracy and phase stability are unchanged for single Faraday components. However, using the smaller restoring beam for multiple Faraday components we find a) better discrimination of the components, b) significant reductions in blending of structures in tomography images, and c) reduction of spurious features in the Faraday spectra and tomography maps. We also discuss the limited accuracy of information on scales comparable to the width of the amplitude spectrum peak, \revv{and note a \emph{clean-bias}, reducing the recovered amplitudes}. 
We present examples using MeerKAT L-band data. We also revisit the maximum width in Faraday depth to which surveys are sensitive, and introduce the variable $W_{max}$, the width for which the power drops by a factor of 2.  We find that most surveys cannot resolve continuous Faraday distributions unless the narrower \emph{full} restoring beam is used.}   
\end{abstract}

\section{Introduction}
% for IEEE journal papers produced under \LaTeX\ using
% IEEEtran.cls version 1.7 and later.
%\IEEEPARstart
The technique of  Faraday Synthesis, introduced by \cite{Burn66} and developed into a formal tool by \cite{Brentjens2005} (hereinafter \bdb), allows one to separate polarized emission coming from regions of differing Faraday depths that are combined in the observed Stokes parameters.  Since almost all optically thin radio sources are depolarized, i.e., their fractional polarization decreases as the wavelength increases, this implies that they have a range of Faraday depths within the solid angle of an individual observing beam. These can result from  variations in the Faraday depth through different individual lines of sight, leading to what is termed ``beam" depolarization, or through the interleaving of Faraday and synchrotron emitting regions along the line of sight, leading to ``internal" depolarization.   In either case, the range of Faraday depths in the polarized emission can be a powerful diagnostic of multiple emitting regions and the associated magnetized thermal plasmas.  

The ability to effectively separate multiple Faraday depths has therefore been a subject of intense interest.  It has led to the deployment of wideband receiving systems and surveys, e.g., ASKAP \citep{Heywood2016}, MeerKAT \citep{Jonas2016}, LOFAR \citep{vanH2013}, VLASS \citep{Lacy2020}, and uGMRT \citep{Suresh2014},  which provide the requisite coverage in $\lambda^2$ space. It has spawned multiple efforts to deconvolve the direct or  ```dirty" Faraday spectrum, to remove the sidelobes arising from the incomplete $\lambda^2$ coverage \citep[e.g.,][]{Heald2009b,Frick2011,Andrecut2012,Ndiritu2021}.   Also, there is a strong interest  in detecting ``complexity" in Faraday spectra, i.e., the presence of more than one Faraday component \citep{Brown2019, Alger2021, Cooray2021, Pratley2021}. A parallel set of efforts have used parametric methods, e.g. Q-U fitting, based on prior knowledge of the form of the Faraday spectrum, \cite[e.g.,][]{Farnsworth2011, Osullivan2012}. \rev{\cite{Sun2015} compare the performance of a variety of different techniques for extracting information about complexity in the Faraday spectra.}

%In this paper, we look at a simple alternative to best exploit the information in the real and imaginary parts of the complex Faraday spectrum. 
\rev{The simplest formulation of Faraday synthesis, as introduced by \cite{Burn66}, used the kernel $e^{2i\phi\lambda^2}$ in the transform, where $\phi$ is the Faraday depth.  \bdb~ introduced a different kernel, $e^{2i\phi(\lambda^2-\lambda_0^2)}$ with $\lambda_0^2 = \langle\lambda^2\rangle$, where $\lambda$ ranges over the observed wavelengths;  the stated goal was to increase the stability of the recovered phases/polarization angles in the deconvolution process. This formulation is widely used today. This smooths out the variations in the complex Faraday beam, as intended,  while leaving the amplitude beam unchanged. During the deconvolution process, the clean components are then restored using the width of the amplitude beam. The question being asked in this paper is whether information is lost through this process, and can be recovered using a restoring beam matched to the narrower width of the main real lobe of the original, \citep{Burn66}, complex spectrum.  Throughout,  the Faraday spectrum produced using $\lambda_0^2 = \langle\lambda^2\rangle$ \rev{will be called} \emph{nominal}, \nomi, while the spectrum produced with no $\lambda_0^2$ (effectively, $\lambda_0^2=0$) will be called \emph{full}, \ful.}  %  This choice minimizes the phase (angle) variations as a function of the observed Faraday depth, $\phi$, relative to its true value.  This occurs at the expense of broadening the complex beam structure; it leads to an effective resolution in Faraday space that we hereinafter call the ``nominal" resolution.   Here, we investigate a different choice, effectively setting $\lambda_0^2=0$, which provides a much narrower \emph{real} beam, which we term ``full" resolution. The width of the beam in the \emph{amplitude} spectrum is independent of the choice of $\lambda_0^2$.  The question we ask is whether additional information can or cannot be derived using the ``full" resolution spectrum.  The answer, as we will show, is that the results of the two methods are essentially identical when there is a single underlying Faraday component, but there are important regions of parameter space where ``full" resolution spectra can better identify complexity in the Faraday spectra, and also reproduce more accurately the spatial distribution of Faraday structures.

\section {Faraday Synthesis - Nominal and Full Resolution\label{synthesis}}
%We start by briefly summarizing the basis of Faraday synthesis, and then introduce the different ways of looking at the resolution, based on the complex nature of the Faraday spectrum.

%A linearly polarized wave passing through a magnetized plasma will
%experience a Faraday rotation of the angle of the polarized signal
%(\bdb):
%\begin{equation}\label{rot}
%\Delta \chi\ =\ \lambda^2\ 0.81 \int n_e B_\parallel dr,
%\end{equation}
%where $\lambda$ is the wavelength in $m$, $n_e$ is the electron density in
%$cm^{-3}$, $B_\parallel$~is the strength of the component of the
%magnetic field along the line of sight in $\mu$Gauss and $r$ is distance
%in parsec.
%The concepts of Faraday dispersion and Faraday depth were introduced by
%\cite{Burn66}. 
%Faraday Synthesis is a technique for examining the Faraday dispersion
%function (spectrum) along a given sight-line introduced by \cite{Burn66} and
%further developed by \bdb.

%\cite{Heald2009}  complex CLEAN, FT Q+iU not (Q+iU)/I?? 
%Use $\lambda_0$ the average lambda, this minimizes the imaginary
%component,  Does this give the lower resolution?

%$\lambda_0$, Brentjens2005 says more stable if weighted average of
%$\lambda$ used.  May be best in RMSyn since only RM (use $\phi$) and
%polarization intensity determined.

%\subsection { Faraday Synthesis in Obit}
Both the full and nominal resolution spectra and deconvolution were implemented in the Obit package
\citep{OBIT} \footnote{http://www.cv.nrao.edu/$\sim$bcotton/Obit.html}.
 This implementation allows usage of Q and U images unequally spaced in frequency, which preserves some of the frequency resolution while meeting other needs, such as more uniform coverage in $\lambda^2$ space. The task MFImage directly transforms the input Q and U data, without normalizing by I.  To accurately recover the Faraday spectrum for a single component, the spectral dependence must be removed, e.g.,  by using Q/I and U/I;  for multiple components \rev{in a single beam}, with potentially different spectra, this may not be possible.   In this paper, we assume only flat spectra \rev{for our simulated signals}. 
%The broadband imaging in task MFImage does a
%joint deconvolution over the entire frequency band while preserving
%some frequency resolution although unequally spaced in frequency.

%Formally, the Faraday spectrum is the Fourier transform of Q+iU in
%$\lambda^2$ space.
%In practice, only a portion of $\lambda^2$ space can be accessed and,
%in particular, the negative portion is completely inaccessible.
%This latter condition represents a fundamental limitation as the
%Faraday spectrum will, in general, be complex and not necessarily
%symmetric so the Q+iU function will not be symmetric about $\lambda^2$
%= 0.

The Faraday spectrum is approximated using the Fourier series
\begin{equation} \label{F_dispersion}
 F_k(x,y)\ =\ K \sum_{j=1}^{n} W_j \ e^{-2 i \phi_k (\lambda_j^2-\lambda_0^2)}
[Q_j(x,y) + i U_j(x,y)] 
\end{equation}
for Faraday depth $\phi_k$ where $W_j$ is the weight for frequency
sub-band $j$ of $n$, $\lambda_j$ is the wavelength of frequency sub-band
$j$, $\lambda_0$ is the reference wavelength, $i$ is $\sqrt{-1}$ and
$Q_j$ and $U_j$ are the Stokes Q and U sub-band images at frequency  $j$. 
The normalization factor $K$ is $1/\sum_{j=1}^{n} W_j$.   

$W_j$  may also include a correction for spectral index\footnote{The
  spectral index is defined as $I_\nu \propto \nu^{\alpha}$.}, 
$\alpha$:
\begin{equation} \label{weight}
 W_j \ = \ w_j e^{-\alpha \log(\nu_j/\nu_0)} 
\end{equation}
where $\nu_j$ is the frequency of channel $j$, $\nu_0$ is the
reference frequency (corresponding to $\lambda_0$) and the weight for sub-band $j$, $w_j$, is
derived from the off--source RMS in the $Q_j$ and $U_j$ images.
$w_j$ is zero for frequency bins totally blanked due to RFI
filtering; \rev{in our simulations the noise is the same in all channels, so $w_j = 1$ for the non-blanked channels. The task RMSyn optionally allows a correction for a default spectral index for an entire image; this was not used in our flat-spectrum simulations.}  

%Faraday synthesis is implemented in Obit task RMSyn which is given Q and U spectral cubes and produces a Faraday spectrum cube on a user defined grid of values. 

 % The broadband imaging in task MFImage does a joint deconvolution over the entire frequency band, using  
%\rev{A joint deconvolution over the entire frequency band is done by OBIT task MFImage;  it works on} a pixel-by-pixel basis using a complex H\"ogbom 
%CLEAN  \citep{Hogbom74} similar in implementation to
%\cite{Heald2009}. 
%The CLEAN proceeds using a user specified loop gain (default 0.1) up
%to a user specified maximum number of iterations and/or a maximum
%residual to collect a set of complex delta functions in bins of $\phi_k$. 
\rev{A deconvolution over the entire frequency band is done by
OBIT task RMSyn; it works on a pixel-by-pixel basis using a
complex Högbom CLEAN \citep{Hogbom74} similar in implementation
to \cite{Heald2009}. The CLEAN proceeds using a user specified
loop gain (default 0.1) up to a user specified maximum number of
iterations and/or a maximum residual to collect a set of complex
delta functions in bins of $\phi_k$}.  The Faraday beam is calculated over twice the extent in $\phi$
  as the Faraday spectrum to allow deconvolution over its full range.\rev{ The complex Faraday spectrum is convolved with a Gaussian restoring beam, the choice of which is described in the next section. }
  
%  In this paper, all spectra and images have undergone deconvolution}
%Such a deconvolution is necessary to maintain high dynamic range in
%complex fields.

\begin{figure}
\centering
   \includegraphics[width=3.5in]{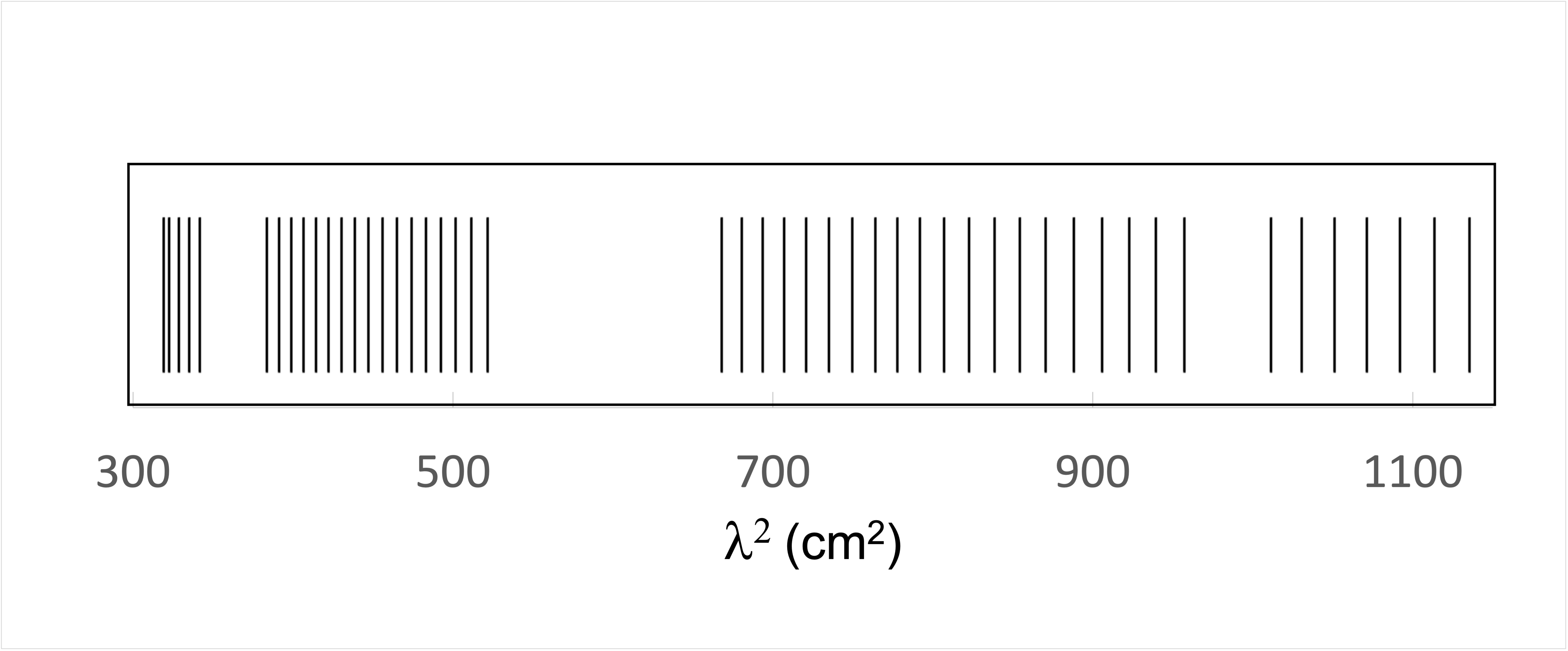}
\vspace{-0.1in}
\caption{Wavelength$^2$ coverage for the simulations presented in this paper. It approximates that available for actual data from MeerKAT L-band, \citep[e.g.,][]{Knowles2021}. Lines denote the central wavelengths for each channel and the coverage is continuous except in the large gaps.}
\label{coverage}
\end{figure}

The following simulations use the  MeerKAT L-band frequency coverage, 68 channels of 1\% (varying) bandwidth, from 890 to 1681 MHz, with channels removed where MeerKAT experienced RFI, so as to more closely approximate realistic data sets. The trimmed data set contained 49 channels. The resulting coverage in $\lambda^2$ space  is shown in Figure \ref{coverage}.

\subsection{\rev{Choice of Reference Wavelength and ``Resolution"}}

\begin{figure*}
\centerline{
    \includegraphics[width=3.5in]{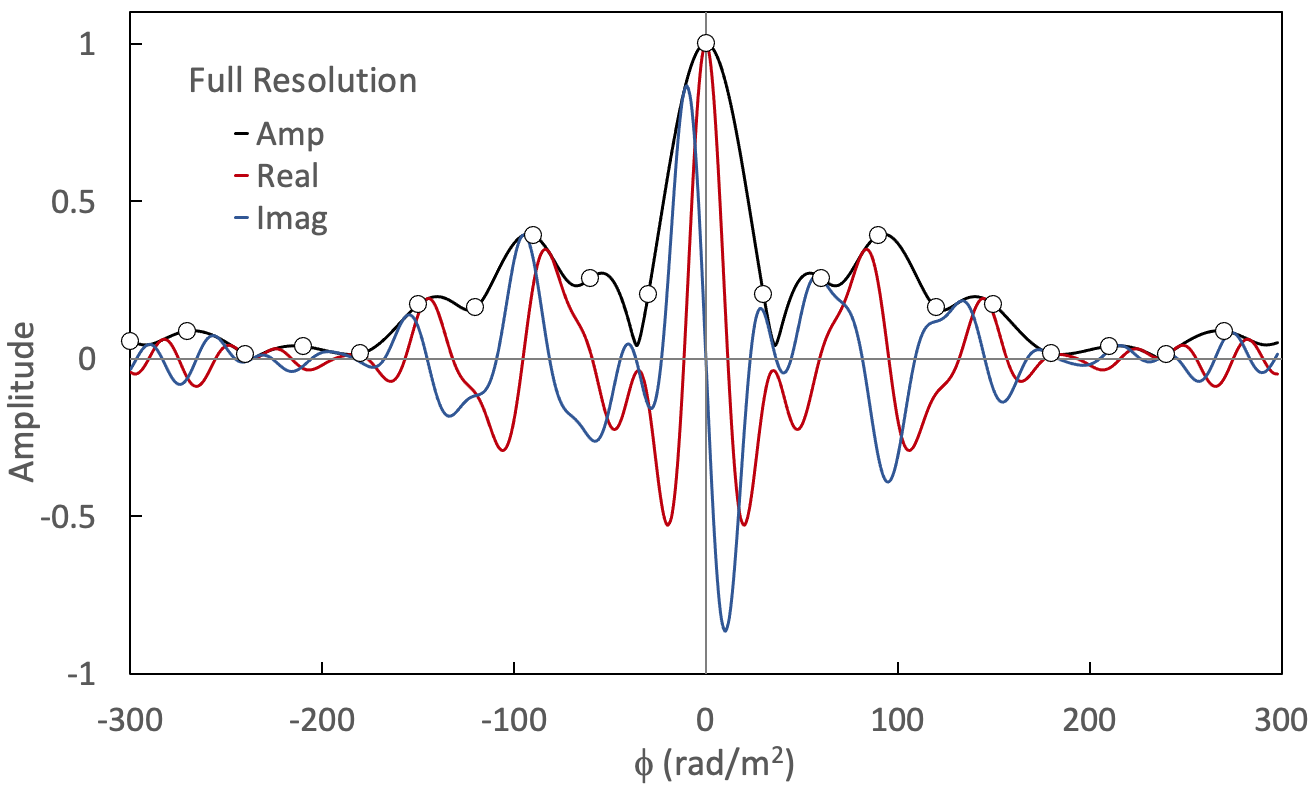}
    \includegraphics[width=3.5in]{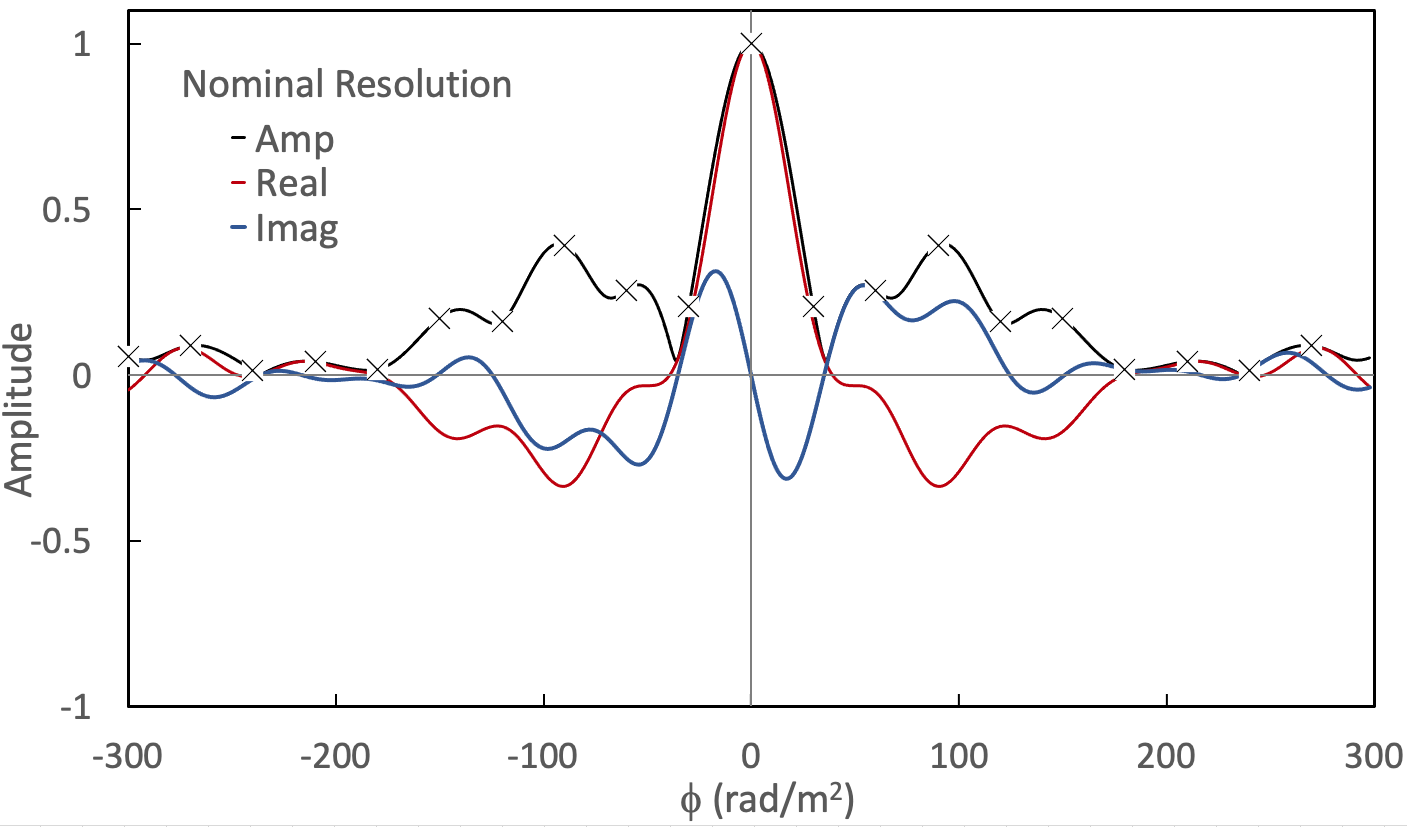}

}
\caption{ The complex Faraday spectra for \ful~ and \nomi~.
The amplitudes, real and imaginary spectra are shown in black, red, and blue, respectively. "X" symbols denote the \nomi~ results, while open circles denote the \ful~ results.  These same symbols will be used where needed in all subsequent figures. Note that the amplitude spectrum is identical for the two methods; only the real and imaginary parts differ.
} 
\label{beam}
\end{figure*}
\rev{We start by looking at the Faraday spectra created for the default case \ful, effectively with $\lambda_0^2=0$, and for the \bdb~ implementation, with $\lambda_0^2=\langle \lambda^2 \rangle$. 
In practice, to avoid numerical problems, we set $\lambda_0^2=10^{-6}$ for \ful, but will refer to this as $\lambda_0^2=0$ hereinafter. For \nomi, we use $\langle\lambda\rangle^2$ (0.065 $m^2$) instead of $\langle\lambda^2\rangle$ (0.067 $m^2$); this very small difference does not affect any of the results. The Faraday beams for these two choices are shown in Figure \ref{beam}. The amplitude beams for the two methods are identical, since only the phases of the complex spectrum have been shifted.  For \nomi, the main lobe of the real beam is considerably broader, by design, since \bdb~ intended to minimize the changes in phase as a function of Faraday depth, $\phi$.  Below, we examine whether this shift of reference wavelength accomplishes its intended purpose. }
%The main lobe of the real component is indeed narrower for \ful, as desired, while the phase of the spectrum changes more rapidly away from the peak. \footnote{Note that for the {\em nominal} case, . This arbitrary choice has no effect on the comparisons presented here.}

\rev{The widths of the main lobes in the  spectra determine both the accuracy of Faraday depth determinations as well as the identification of complex structure (i.e., emission at more than a single Faraday depth).  However, we specifically avoid using the term ``resolution" at this point, because it presumes that we have established how the performance depends on the widths of the amplitude and real beams. Instead, we give empirically-based names to these widths, namely:}
\begin{equation}
    \Phi_{nom} \equiv FWHM~real~peak~for~ \lambda_0^2~ =~ \langle\lambda\rangle^2, and
    \end{equation} 
    \begin{equation}
    \Phi_{full}\equiv FWHM~real~peak~for~ \lambda_0^2~ = 0
\end{equation} 
%\rev{We note that for \nomi, the FWHM of the amplitude peak $\Phi_{amp}= 1.09\times \Phi_{nom}$, so this distinction is not significant for our purposes.  In the task MFImage, there is the option to use restoring beams which are fit to the main lobe, or an arbitrary restoring beam can be specified.}

%\textcolor{red}{move this}
%The reference wavelength in Equation \ref{F_dispersion}, $\lambda_0$. is relatively arbitrary.
%\bdb~ reported that  solutions were more
%stable when $\lambda_0$ was in the bandpass observed, e.g. the
%weighted average of the sampled wavelengths.
%This was likely due to the rapid fringing inside the amplitude
%envelope in their Figure 3 when $\lambda_0=0$.
%Subsequent implementations of Faraday synthesis have followed this
%practice of choosing $\lambda_0$ in the middle of the band. 

%As shown below, we found that using the MeerKAT bandpass, this choice did not improve the stability. Further study would be useful to determine whether stability depends on a relatively small number of fringes within the amplitude beam when $\lambda_0=0$.  This number is proportional to proportional to $\frac{\lambda_{max}^2+\lambda_{min}^2}{\lambda_{max}^2-\lambda_{min}^2}$, which for MeerKAT L-band is 1.8, while it was 7.5 for \bdb's WSRT 92~cm data.

 %\rev{with a width either fit to the real component of the spectrum  or specified by the user. }
  
\rev{ The Faraday amplitude spectrum is calculated from its convolved real and imaginary parts and is identical for \ful~ and \nomi. For \nomi, the widths of the real beam and the amplitude beam are almost the same, by construction.  For \ful, the real beam has a much narrower main peak, and using this narrower width for the restoring beam forms the basis for the experiments in this paper.  See Figure \ref{beam}.  }

%\begin{table}[]
%\vskip 0.25in
%\caption{Resolution in Faraday space}
%\vskip 0.1in
%\begin{center}
%\begin{tabular}{|l|c|}   
%\hline
%\hline
%\vspace{0.1in}
%  Nominal & $\frac{\sqrt(3)}{\lambda_{max}^2-\lambda_{min}^2}$\\
%  \hline\
%  \vspace{0.1in}
%  Full & $\frac{\sqrt(3)}{\lambda_{max}^2+\lambda_{min}^2}$\\
%%\hline
%\end{tabular}
%\end{center}
%\hfill\break
%\label{resolution}
%\end{table}

%The full width at half maximum of the main peak in the Faraday beam determines the resolution in Faraday space. We are testing here whether the use of the width of the {\em real} component only, which depends on the choice of $\lambda_0$, affects the accuracy of depth measurements or complexity. For clarity, we therefore adopt the following nomenclature:

  The exact values of $\Phi$ depend on the details of coverage in $\lambda^2$ space, gaps in coverage, any weighting, etc.  Approximate expressions are very useful, however, and for \nomi~ is given by \cite{Dickey2019} as:
 \begin{equation} 
   \Phi_{nom} \approx \frac{3.8}{\lambda_{max}^2 - \lambda_{min}^2}.
   \label{phinom}
 \end{equation}
 \rev{since} for \nomi, \rev{where the widths of the real and amplitude peaks are almost the same (see Figure \ref{beam})}, then $\Phi_{nom} \sim \delta\phi$, as given by \cite{Dickey2019}.
 
 To approximate the value of $\Phi_{full}$, we note that using the integral version of Eq. \ref{F_dispersion}, zero values in $\mathbb{R}$(~\ful~) occur when 
 \begin{equation}
 \revv{    2\phi\lambda_{max}^2 = 2\phi\lambda_{min}^2, ~ ~ ~ ~ and}
 \end{equation}
 \begin{equation}
  \revv{   2\phi\lambda_{max}^2 = \pi -  2\phi\lambda_{min}^2 }
 \end{equation}
 %Because a sine wave is symmetric around $\pi/2$, the first zero occurs when, for some value $\delta$, \begin{equation}
 %    2\phi\lambda_{max}^2 = \pi/2 - \delta,
% \end{equation} 
 %and, simultaneously 
% \begin{equation}
 %    2\phi\lambda_{min}^2 = \pi/2 + \delta.
% \end{equation}  
\revv{ The first condition ($\lambda_{max}= \lambda_{min}$) is not meaningful, and the second condition is satisfied for}
 \begin{equation}
     \phi=\frac{\pi}{2(\lambda_{max}^2 + \lambda_{min}^2)}.
 \end{equation}
 The FWHM of one half-cycle of a sine wave occurs at approximately the value of the first zero crossing.  Fitting a Gaussian to the main lobe of yields a slightly different value, which we adopt here, of 
\begin{equation} 
  \Phi_{full} \approx  \frac{2}{\lambda_{max}^2 + \lambda_{min}^2}.
\label{phiful}
\end{equation}

 % It is important to note that $\Phi_{full}$ is only weakly
%  dependent on $\lambda_{min}$, since $\lambda_{max}^2+\lambda_{min}^2 \sim \lambda_{max}^2$
 % for a wideband system.  %In the limit where $\lambda_{min}\approx \lambda_{max}$, there is only a single Fourier component in the spectrum, and the real beam becomes a sine wave (\ful) or a constant (\nomi).  
  
  We fit Gaussians to the central peak in the real spectra (Figure \ref{beam}). We find $\Phi_{nom} = 45$~\radmm~ and $\Phi_{full} = 16$~\radmm~, similar to the values calculated from Eqs. \ref{phinom} and \ref{phiful} of 42 and 14 \radmm, respectively.

%  Using a near zero reference wavelength leads to a higher phase
%gradient across the FPSF functions and a narrower central real
%lobe. %, hence better resolution in $\phi$.
%The MeerKAT L band frequency coverage with zero $\lambda_0$
%illustrated in Figure \ref{DirtyBeam} left contains only a single real
%lobe inside the overall envelope so has minimal opportunity for lobe
%ambiguity while it still has a narrower central real lobe than using a
%mid--band $\lambda_0$. 
%The following development uses a reference wavelength close to
%zero\footnote{But not exactly zero to avoid numerical problems.}
%and the width of the inner real lobe as the width of the restoring
%function. 

\begin{table*}
\vskip 0.25in
\caption{Faraday spectrum parameters}
\vskip 0.1in
\begin{center}
\begin{tabular}{|l|c|c|c|c|c|c|c|}   
\hline
\hline
  Survey & Freq.  & Wavelength$^2$ & Nom. Res. & Full Res. & Ratio  & \rev{$W_{max}$} & $\frac{\lambda_{max}}{\lambda_{min}}$ \\
   & MHz & cm$^2$ & \radmm & \radmm & (Nom/Full) & \radmm &\\
\hline
MeerKAT L-band & 900 - 1600  & 318 - 1135 & 45* & 16* & 2.8* & \rev{25} & 1.9\\
\hline
POSSUM Band 1   &800 - 1088     &760 - 1406     & 59   &9  &6.4 &\rev{14}  &1.4\\
\hline
POSSUM Band 2  &1152 - 1440 & 434 - 678&156 &18 &8.7 &\rev{25}  &1.3\\
\hline
VLASS  &2000 - 4000     &56- 225     &225   & 71    &3.2    & \rev{148}   &2.0\\
\hline
LOFAR (HBA)  &120 - 240 &15625- 62500 & 0.8  & 0.3 &  3.2 & \rev{0.5} &  2.0 \\
\hline
uGMRT Band 3 &250 - 500     & 3600 - 14400  & 3.5  & 1.1  & 3.2  & \rev{ 2}  &2.0\\
\hline
uGMRT Band 4 &550 - 850     & 1146 - 2975  & 22.0  &4.7   & 4.6  &\rev{8}   &1.5\\
\hline
Apertif  &1130 - 1430  & 440 - 705  &144   & 17.5  & 8.2  & \rev{25}  &1.3\\
\hline
WSRT (92cm) & 319-365   &6560-9025  & 15.2  & 1.3&11    &\rev{2}  & 1.18\\
\hline
SKA1 Low &50 - 350  & 7347 - 36000  & 0.11  & 0.05   &2.0   & \rev{0.9}  &7.0\\
\hline
SKA1 Mid 1 &350 - 1050 &816 - 7347 &5.8 &2.5 & 2.4&\rev{9} &3.0\\
\hline
SKA1 Mid 2  &950 - 1760 &291 - 997 &53.8 &15.5 &3.5 &\rev{30} &1.9\\
\hline
SKA1 Mid 3 &1650 - 3050 &97 - 331 & 163& 47& 3.5& \rev{89} &1.8\\
\hline
\end{tabular}
\end{center}
Note:  * indicates measured values for the MeerKAT coverage discussed here.  All other Faraday spectrum parameters use the approximate calculations described in the text.
\hfill\break
\label{surveys}
\end{table*}
Table \ref{surveys} summarizes the Faraday spectrum parameters for a variety of telescope/receivers/surveys that are used for polarization studies.  \rev{One key parameter is  $\rho = \frac{\lambda_{max}}{\lambda_{min}}$, which can affect the impact of using $\Phi_{full}$ and the ability to resolve complex structures.   We can write 
\begin{equation}
\frac{\Phi_{nom}}{\Phi_{full}}~ =~ 1.9~ \frac{\rho^2+1}{\rho^2-1}
\end{equation}
For very wide bands, where $\rho >> 1$, the reduction in width using $\Phi_{full}$ is $\approx$1.9.   For narrow band observations, where $\rho$ approaches 1, $\Phi_{full}$ approaches a constant value \revv{$\approx\lambda_{max}^{-2}=\lambda_{min}^{-2}$} while $\Phi_{nom}$ becomes very large. The relative performance of \ful~ and \nomi~ shown in this paper is for the case of $\rho~\sim~2$, similar to other wideband surveys;  whether the results are applicable to much narrower band observations, such as for Apertif and  the WSRT 92cm studies of \bdb, would need further study.}  

\rev{Of particular interest is whether a survey can  have sensitivity to a continuous distribution of Faraday depths -- quantified here by a new parameter, \emph{maximum-width},  $W_{max}\approx~ 0.67*\lambda_{min}^{-2}(1~+~\rho^{-2})$~ \footnote{\rev{Previously, the quantity \emph{max-scale}, as defined by \bdb, was used to address this issue.  In the Appendix, we show that $max-scale~=~\pi~\lambda_{min}^{-2}$  significantly overestimates the sensitivity to broad Faraday depth distributions.}}, and simultaneously have a sufficiently narrow Faraday beam to resolve the structure.  As we will derive in the Appendix, this simultaneous condition is marginally met in most cases when $\Phi_{full}$ is used, but for $\Phi_{nom}$, this requirement is satisfied only for $\rho~>~2.4$; this is true only for  SKA1-Mid and SKA1-Low in this survey compilation. Thus, none of the other surveys will be able to resolve continuous Faraday structures if $\Phi_{nom}$ is used.}  %There is also a ''maximum scale," i.e., the width of the polarized emission in Faraday depth space, beyond which the amplitude in the Faraday spectrum falls by more than a factor of 2, and eventually become non-detectable.  \bdb quantify this scale as $\phi_{max} \sim \frac{2\sqrt(3)}{\delta\lambda^2}$, where $\delta\lambda$ is the width of an individual channel. Thus, structures that are too broad in Faraday space The ability to simultaneously detect and resolve complex structures thus depends on the ratio of the maximum scale to the resolution.  This varies substantially in different surveys;  it is approximately measured by the $\frac{\lambda_{max}}{\lambda_{min}}$ ratio given in the final column. In most cases, the ability to resolve complex structure is marginal.  This will become obvious in Section \ref{tophat}. The ratio $\frac{\Phi_{full}}{\Phi_{nom}}$ also varies significantly across surveys.  It is not clear whether our results where that ratio is small apply to Apertif, or POSSUM Band 2, or WSRT 92cm, with much larger values;  this is an area for further study.

\begin{figure}
\centering
\includegraphics[width=3.1in]{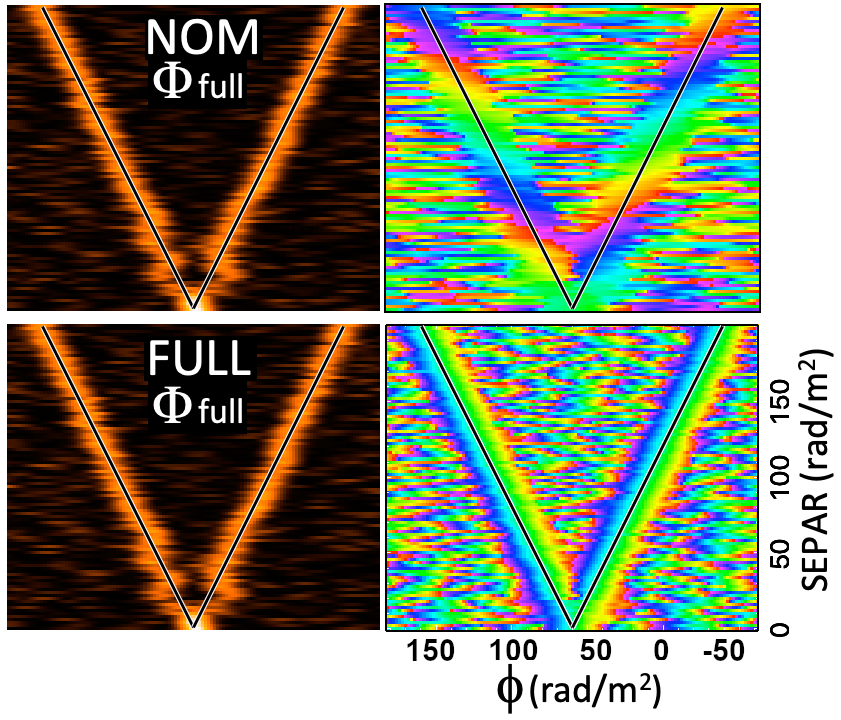}
\label{equiv}
\caption{\rev{Faraday spectra, one in each row, for a series of simulated signals. Each simulation includes two input delta-function Faraday components at various separations, indicated by the black lines, with the addition of random noise in each $\lambda^2$ channel. The same noise is used for both \ful~ and \nomi.   Results are shown for both,  \emph{but using the same restoring beam corresponding to $\Phi_{full}=16$~\radmm}.  Top frames show the amplitude and phase for \nomi, spectra, while bottom frames show the same for \ful. By default, for \ful, the phases are those at 0 wavelength, while for \nomi, the phases are at $\lambda_0$. }\revv{Note that even for \ful, there are phase gradients across the spectra of each component that arise during the clean process. This will become relevant in other experiments described below.} }
\end{figure}
  
  \rev{Before comparing results with these two methods, and their two different restoring beams,  we compare their outputs with a single fixed restoring beam. If we start with cubes of $Q(\lambda^2), U(\lambda^2)$, the amplitude cubes \ful~ and \nomi~ produced from them will be identical, prior to deconvolution. The phases, and thus the real and imaginary spectra, will differ, since \ful~ phases are at \revv{$\lambda_0 = 0$}, and \nomi~ phases are at \revv{$\lambda_0 =\langle \lambda \rangle$}. We performed a variety of different experiments with simulated Q and U distributions similar to those in the following sections where they are described in more detail; the experiments included pure signals and ones with added random noise.  We found that, after deconvolution, the results were still identical for \ful~ and \nomi, \emph{as long as the same restoring beam was used}. The results of one such test are shown in Figure 3.  The top frames show the Faraday amplitude and phase spectra for \nomi, with a different experiment in each row, while the bottom frames show the corresponding results for \ful.  The amplitudes are identical, within rounding errors, as are the phases, after de-rotation for \nomi~ by $2 \phi \lambda_0^2$.  \emph{ Thus, the comparison between \nomi~ and \ful, as discussed in the rest of the paper, is equivalent to a comparison of only the different restoring beams which are used, and not the choice of $\lambda_0$.}}

\section{Recovery of single Faraday components}\label{single}
{\em {Key Findings:}  \nomi~  and \ful~  synthesis\rev{/beams} produce nearly identical results in the recovery of the Faraday depth, polarization angle and amplitude of a single $\delta$ component. Despite the smaller real beam, there is no increased accuracy for \ful.  \rev{At the same time,} the benefits of \rev{the suggested} phase stability for \nomi~ do not result in increased accuracy for the polarization angles.}
\vspace{0.15in}

\begin{figure}
    \centering
    \includegraphics[width=3.5in]{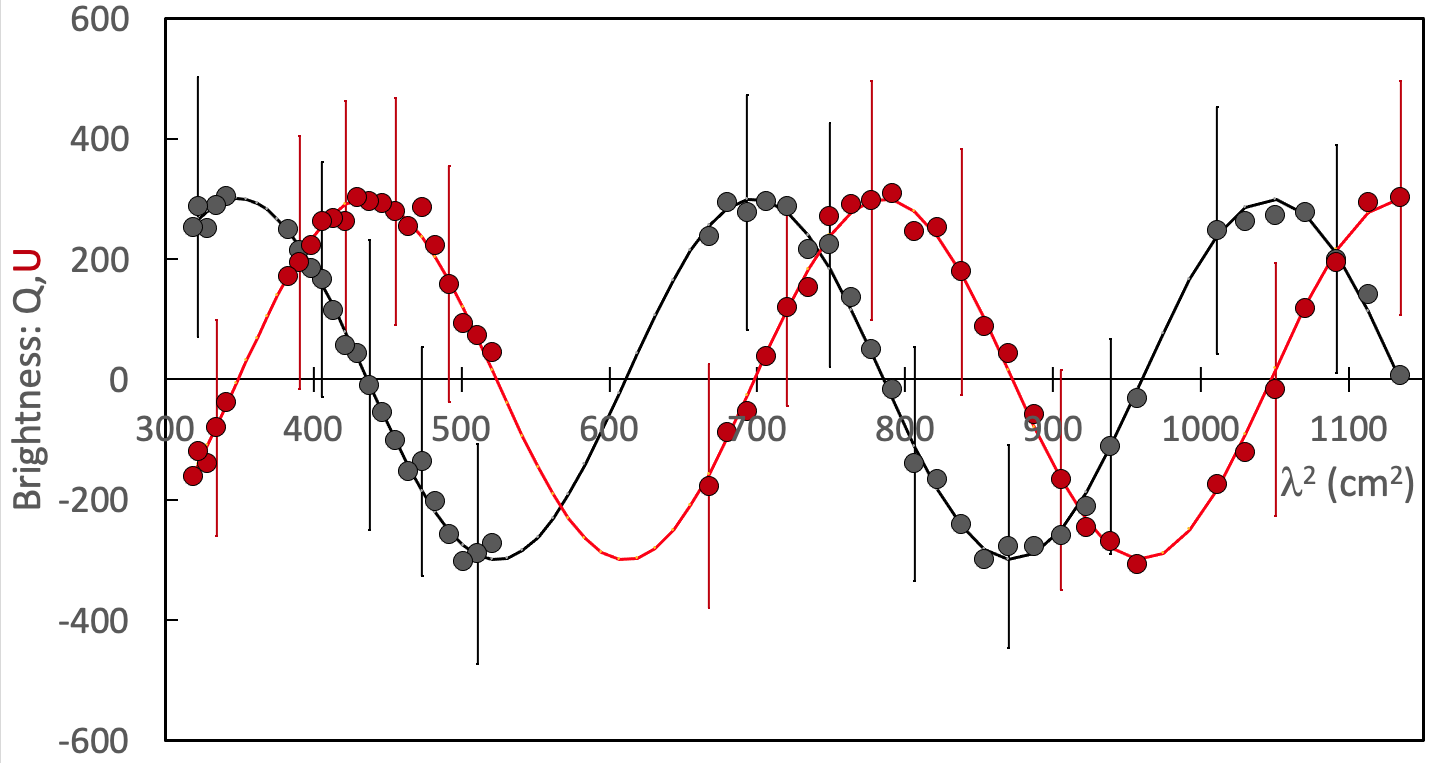}
    \caption{Q, U data \rev{in black and red, respectively,} used for the single Faraday depth experiment at depth $\phi$=90~\radmm.  \rev{Solid curves show the input model, and} each point indicates the average over 101 realizations.  Vertical bars indicate the rms scatter in Q and U among the 101 realizations. }
    \label{oneQU}
\end{figure}

%\subsection{Performance using full resolution}
\rev{Hereinafter, all results from \ful~ (\nomi) use $\Phi_{full}$ ($\Phi_{nom}$) for their respective restoring beams.} Our first test was to compare the ability of \ful~ and \nomi~ resolution Faraday spectra to recover the true parameters of a polarized signal with a single Faraday depth, in the presence of noise.  To that end we simulated signals with Faraday depths ranging from 60~\radmm~ to 160~\radmm.  For each depth, we \rev{added noise to} Q and U at each sampled frequency, for 101 different realizations of the noise (see Figure \ref{oneQU}). The signal:noise per frequency channel was very low, 1.5 \rev{in each realization}.   With 49 frequency channels, the expected signal:noise in the Faraday spectrum was 10.5 .
\begin{figure}
    \centering
 \includegraphics[width=3in]{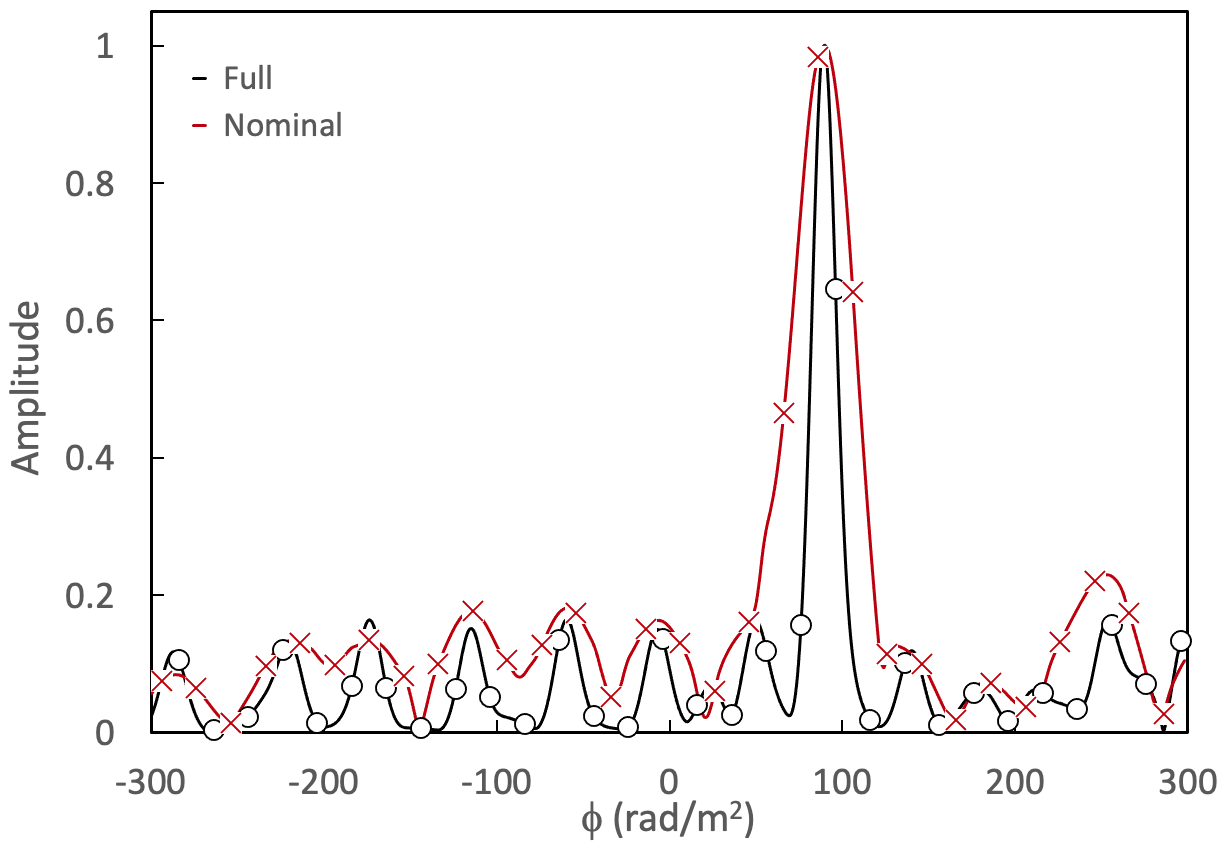}
    \caption{\ful~ and \nomi~ deconvolved, restored Faraday spectra for one realization at depth $\phi$=90~\radmm. The peaks have been normalized to unity. Here and throughout, $\circ$'s circles are used for \ful~ and X's for \nomi.}
    \label{oneQUF}
\end{figure}

 The cleaned, restored Faraday spectra for a single realization at Faraday depth $\phi = 90~$\radmm~ are shown in Figure \ref{oneQUF}. The observed signal:noise values (peak over mean off-peak) were 9.7 for \nomi, almost exactly as expected,  and 17.6 for \ful. The \nomi~ \rev{amplitude} spectrum also appears to be a (not quite perfect) convolution of the \ful~ \rev{amplitude} spectrum.  \rev{Some} differences occur because the convolution by the broader beam happens in the complex space, not in the amplitude spectrum.  %The higher apparent signal:noise in \ful~ is somewhat artificial, since all of the signal here is in the real component, and the real and amplitude beams are very different in this case.  
 Below, we will examine how this \emph{apparent} higher signal:noise and especially the narrower FWHM ($\Phi$),  translates into the uncertainty $\sigma_{\phi}$, \rev{as well as the amplitude and phase of the recovered signals.}  The expected uncertainty \rev{in $\phi$} is given by  
 \begin{equation} \label{pos_scale}
 \sigma\phi\ =\ \Phi/(2 \times SNR)
\end{equation}
where $\Phi$ is the full width at half maximum of the \rev{actual} resolution
and $SNR$ the \rev{actual} signal to noise ratio.

\begin{figure*}
    \centering
    \includegraphics[height=2.1in]{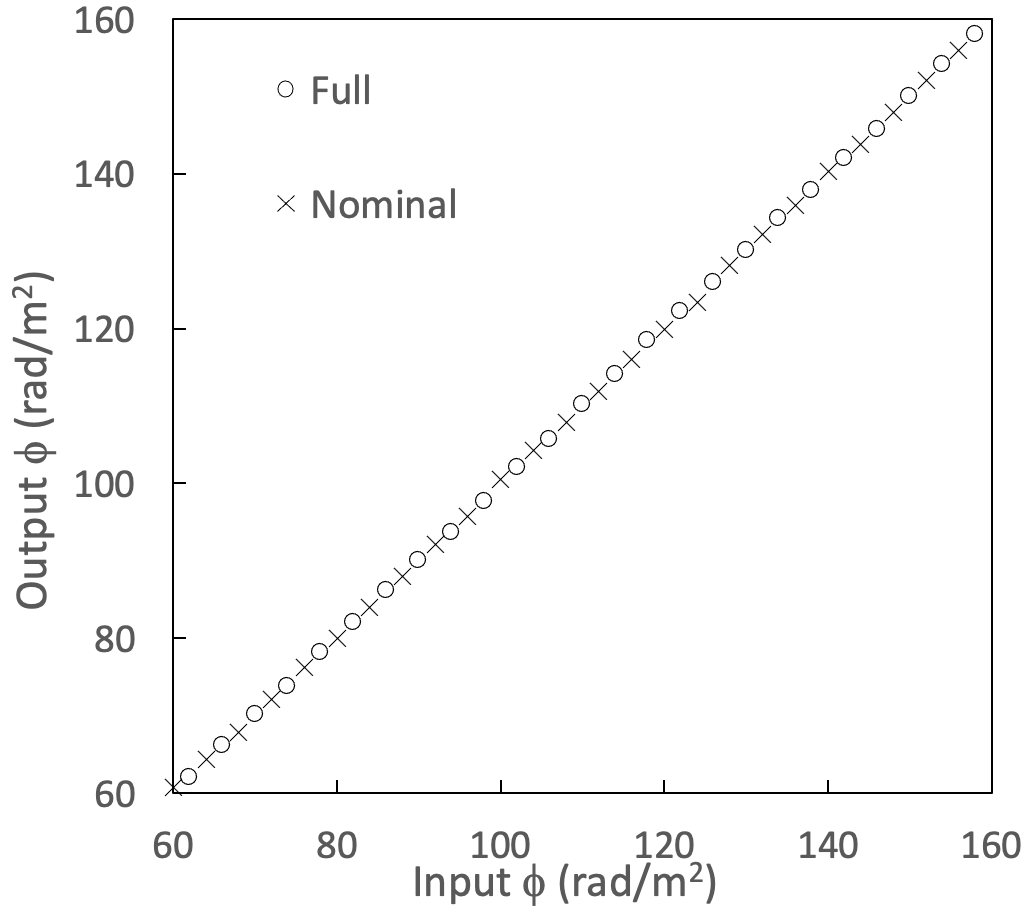}
    \includegraphics[height=2.1in]{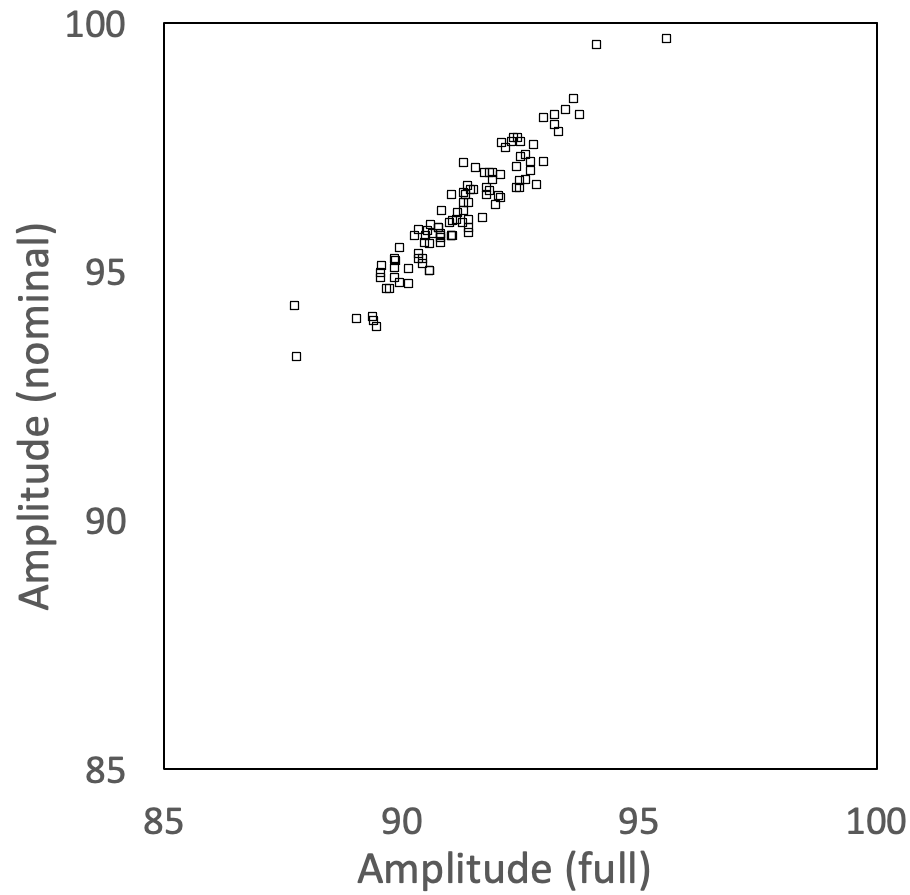}
    \includegraphics[height=2.1in]{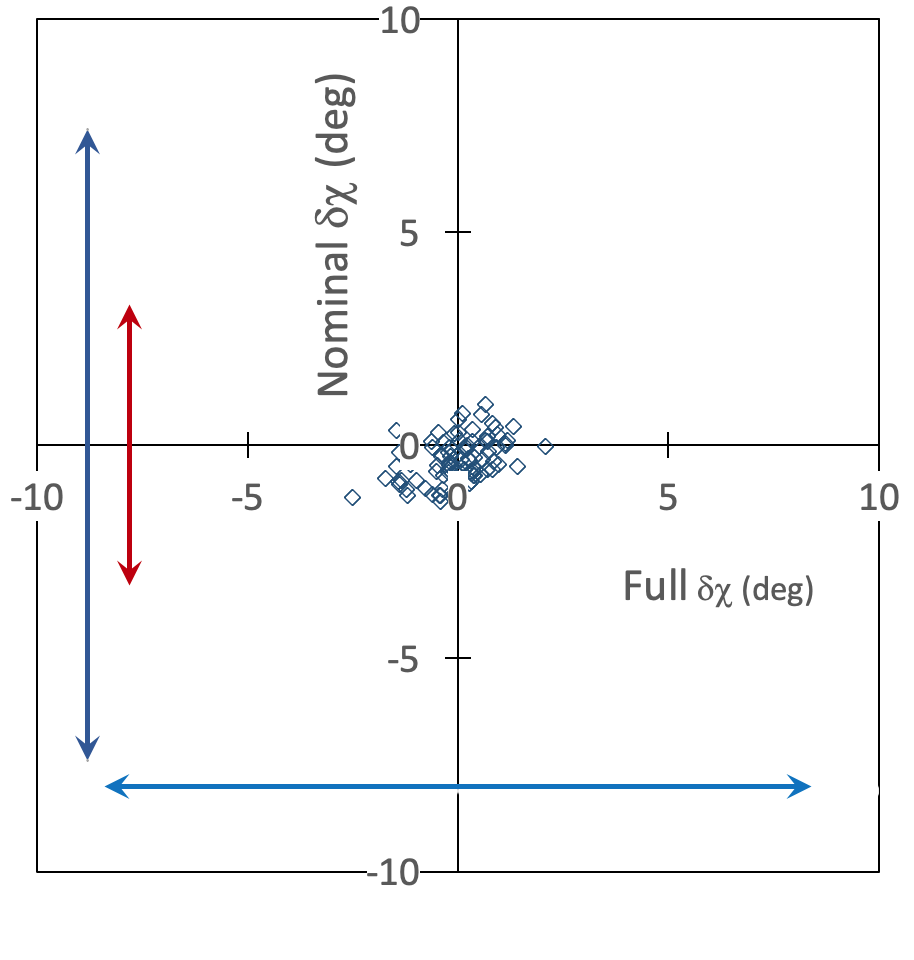}
    \caption{Left: Observed Faraday depths, as a function of input Faraday depth, each averaged over 101 noise realizations. Center: Average amplitude recovered at each Faraday depth, comparing results from \nomi~ and \ful. Right: Average polarization angle $\chi_0$ recovered at each Faraday depth, comparing results from \nomi~ \rev{correcting back to $\lambda=0$ based on the \emph{input} depth} and \ful. The rms scatter in $\chi_0$ \rev{among the 101 realizations} at each Faraday depth is shown \rev{as the black horizontal line for \ful~ and vertical lines showing two different calculations of $\chi_0$ for \nomi, as described in the text.} }
    \label{oneRM}
\end{figure*}

 We compare the results from the two methods over \rev{a large} range of input depths.  Figure \ref{oneRM} shows the observed Faraday depth as a function of input depth, averaged over the 101 realizations at each depth.  
 \rev{The results show that the peak locations of $\phi$ are practically identical for the two methods, as was expected, but had not been previously demonstrated.} %he locations of the peak Faraday depths appear identical with the 2~\radmm~ sampling of the spectrum.
  %We expected the two methods to generate very similar results, in terms of the peak Faraday depth and the amplitude at that depth, as was observed.  
 %It \rev{had not been previously demonstrated} whether the scatter in Faraday depth would be smaller for \ful, since $\Phi_{full} << \Phi_{nom}$ or \rev{whether the opposite would be true,} if the superior phase stability of the \nomi~ spectrum \rev{suggested by \bdb~ } would result in smaller errors in \nomi~ for the observed polarization angle $\chi_0$. % (i.e., at $\lambda=0$) in \nomi.
 Looking now \rev{more closely} at the scatter in the recovered depths \rev{among the 101 realizations at each depth}, Eq. \ref{pos_scale} predicts $\sigma_{\phi}$=2.3 (0.45)~\revv{\radmm}~ for \nomi~(\ful). The \rev{observed scatter} was 2.1 (2.0) \radmm~ for \nomi~(\ful), \rev{on average}.  This is the first important result -- the two methods generate the {\bf same} error in measuring the Faraday depth in the presence of noise. The fact that $\Phi_{full} << \Phi_{nom}$ does \emph{not} improve the accuracy $\sigma_{\phi}$. %Therefore  Eq. \ref{pos_scale} \rev{must use $\Phi_{nom}$ and not $\Phi_{full}$.} %\emph{cannot} use the observed $\Phi_{full}$ and observed signal:noise in the \ful~ spectrum.   

The recovered amplitudes are also well-correlated between the two methods, although their average values differ, as shown in Figure \ref{oneRM}.
%\begin{figure}
%    \centering
%    \includegraphics[width=3.5in]{figslr/One_amps.png}
%    \caption{Average amplitude recovered at each Faraday depth, comparing results from \nomi and \ful. }
%    \label{fig:ampvamp}
%\end{figure}
On average, the \nomi~ amplitudes are $\approx4\%$ low, while the \ful~ amplitudes are $\approx10\%$ low. \revv{By looking at the amplitudes in the dirty spectra, we have verified that this is a \emph{clean bias}, as described by \cite{1998AJ....115.1693C}, where spurious clean components on the sidelobes, whether positive or negative, reduce the amplitude of the peak.  This bias is on the order of the rms scatter, here in the Faraday spectrum, and depends on the amplitude of the sidelobes and the depth of cleaning.  This bias will require correction in any catalogs created using cleaned Faraday spectra. Since the amplitude of the correction depends on the details of the data and processing, simulations will be required in each project. }%   This is related to the SNR, since repeating the experiment at 100:1 SNR resulted in full recovery of the input signal.  Reductions in the Faraday peak amplitude have not been previously documented,  \rev{and  may be related to the power which is scattered into the sidelobe structures}. % The finite sampling in Faraday depth is not responsible, since it creates a much smaller reduction than observed here.

Finally, we turned to the recovered values for the polarization angle $\chi_0$.  Figure \ref{oneRM} plots the average values of \rev{the error in $\chi_0$}, $\delta\chi_0$, \rev{and the rms scatter} for the 101 realizations at each Faraday depth ($\delta\chi_0 = \chi_0$ since the input $\chi_0=0$).  %For \nomi, $\chi_0$ was \rev{first} rotated back to 0 wavelength using the {\em input} Faraday depth. 
\rev{The key question is whether the rms scatter in $\chi_0$ among the 101 realizations at each depth is smaller for \nomi~ as a result of the improved phase stability suggested by \bdb }.  For \ful, \rev{the scatter} is 8.4~\radmm.  For \nomi, we calculated the rms scatter in two different ways.  First, we rotated $\chi_0$ back to zero wavelength assuming the {\em input} Faraday depth, as was done for the averaged points.  This results in an rms scatter of 3.3 degrees; \rev{this is} much less than for \ful, and is the phase stability claimed by \bdb.

However, in an actual experiment, the true Faraday depth would not be known, so we did a second rms calculation based on correcting the observed values of $\chi_0$ to zero wavelength using the {\em observed} Faraday depth.  This results in a scatter of 7.4 degrees, close to the rms scatter for \ful.  We thus conclude that \nomi, despite its superior phase stability at a wavelength of $\lambda_0$, offers little or no advantage in the accuracy with which $\chi_0$ can be recovered at $\lambda=0$.   As mentioned earlier, this conclusion for the MeerKAT bandpass \rev{would need to be verified for narrowband} surveys that have a much larger ratio of $\Phi_{nom}/\Phi_{full}$ (see Table \ref{surveys}).
%\begin{figure}
%    \centering
%    \includegraphics[width=3.4in]{figslr/One_dChi.png}
%    \caption{Average polarization angle $\chi$ recovered at each Faraday depth, comparing results from \nomi~ and \ful. The rms scatter in $\chi$ at each Faraday depth is shown by the horizontal and vertical lines, with a fuller explanation in the text. }
%    \label{fig:dChi}
%\end{figure}

%This small difference could be due largely to the narrower restoring
%FPSF used in the $\lambda_0\sim 0$ test.
%To test this possibility, the $\lambda_0\sim 0$ test was rerun but
%using a restoring FPSF the width of the $\lambda_0$ at midband test.
%The results are shown as the last three entries in Table
%\ref{noiseResTest}. 
%These values for the uncertainty in Peak $\phi$ are very close to
%those for the  $\lambda_0$ at midband test.
%Usage of $\lambda_0\sim 0$ and a restoring FPSF the width of the inner
%real lobe apears not to provide any additional resolution.

\section{Recovery of complex Faraday structure}\label{complex}
In this section, we  examine the ability of nominal and full resolution Faraday synthesis to detect the presence of complex Faraday structure, i.e., where more than one isolated Faraday component is present. The most important regimes are those with Faraday structure on the order of the nominal resolution.  We examine two limiting, idealized noise-free cases:  a continuous distribution of polarized components with constant amplitude, extending over a finite width in Faraday depth \rev{i.e., a `tophat' distribution,} and the more restricted example of two Faraday components separated in depth by $(<1 - \approx5) \times$ the nominal resolution. 
\subsection{Continuous distributions\label{tophat}}
{\em {Key findings:} For a constant amplitude distribution of Faraday depths, at small widths the Faraday spectrum appears as a broadened Gaussian, and at large widths, as two Faraday peaks at the edges.    \ful~ is $\approx 2 \times$ more powerful than \nomi~ in its ability to detect the input Faraday complexity for widths up to at least $\Phi_{full}$. For \ful~(but not for \nomi~), suggestions of the \revv{shape of the} input distribution are apparent in the spectra for widths from $\approx 0.6 - 1.5 \times$ $\Phi_{full}$.}
\vspace{0.15in}

We approximate a continuous Faraday distribution with constant amplitude in Faraday depth (a ``tophat") by summing in Q and U a series individual Faraday delta functions separated by 0.25~\radmm. The amplitudes of the components are normalized so that total input signal flux is constant independent of the tophat width.  We consider cases both where the polarization angle is constant across all components, and also where a gradient in angle is incorporated. 

\begin{figure}
\centerline{
  \includegraphics[width=1.75in]{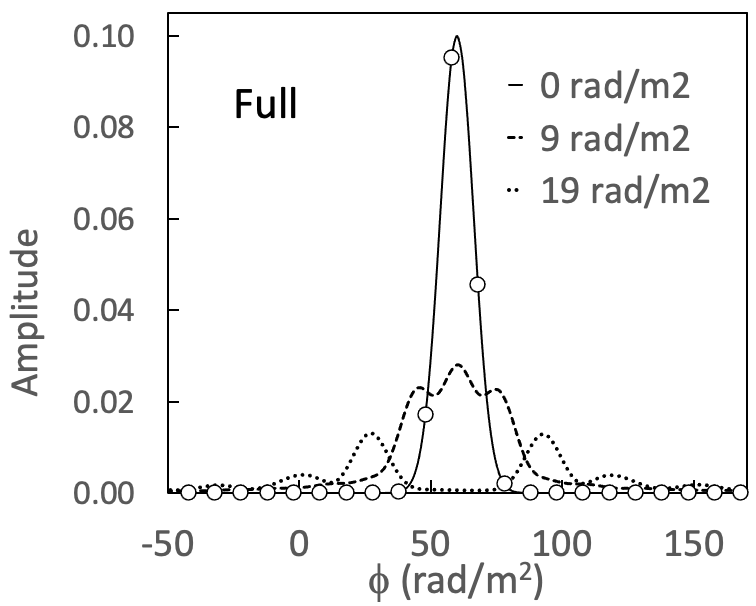}
  \includegraphics[width=1.715in]{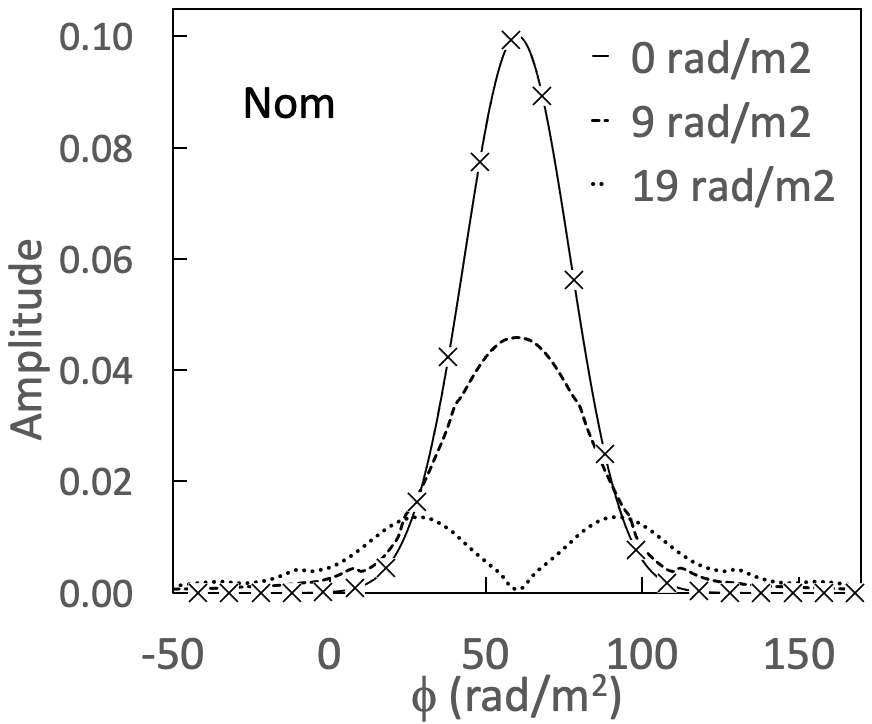}
%  \vskip
%\includegraphics[width=3.5in]{figslr/TopH_slice_Full.png}
}
\caption{Faraday spectra of noiseless simulations of tophat functions
  of various widths in Faraday depth, with a constant polarization angle. Left: \ful; Right: \nomi. \rev{At 9~\radmm~ there is a suggestion of the tophat shape for \ful, but not as clearly for \nomi.}
}
\label{topslice}
\end{figure}
Figure \ref{topslice} shows the resulting Faraday spectra for several different widths, centered on a Faraday depth of 60~\radmm.  The immediate realization is that there is {\em no} width at which a clear tophat shape is seen for \nomi. When the width is too small, the Faraday spectrum appears as a broadened Gaussian.  When the width is too large, the tophat structure is lost, and is replaced by a pair of narrow components at the edges of the distribution.  This behavior was identified by \bdb, who noted (in their Eq. 64) that in order to resolve complex structure, the Faraday resolution must be  less than the maximum  \rev{width where the signal can be recovered. This is discussed further, below. }

\begin{figure}
\centering
   \includegraphics[width=3.5in]{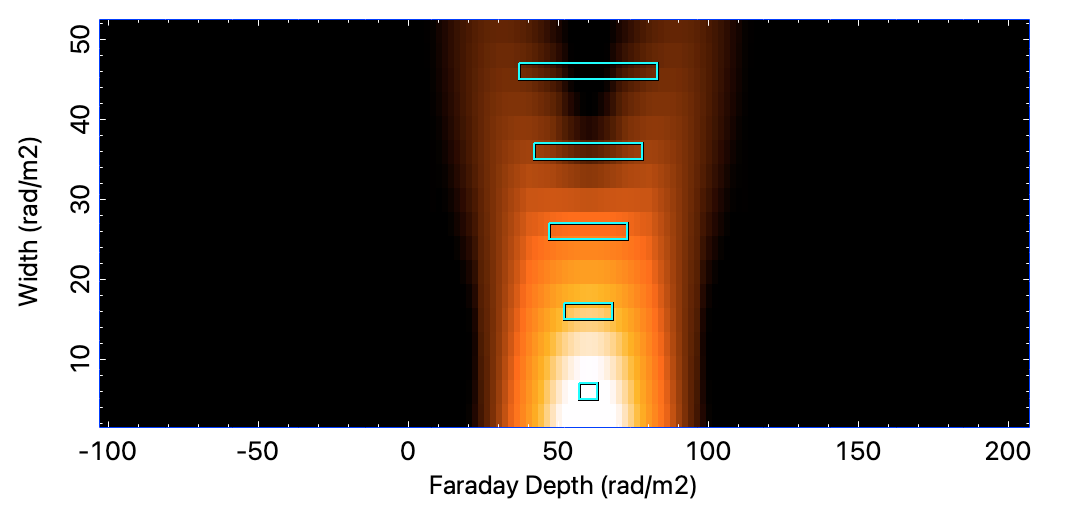}
   \vspace{0.01in}
      \includegraphics[width=3.5in]{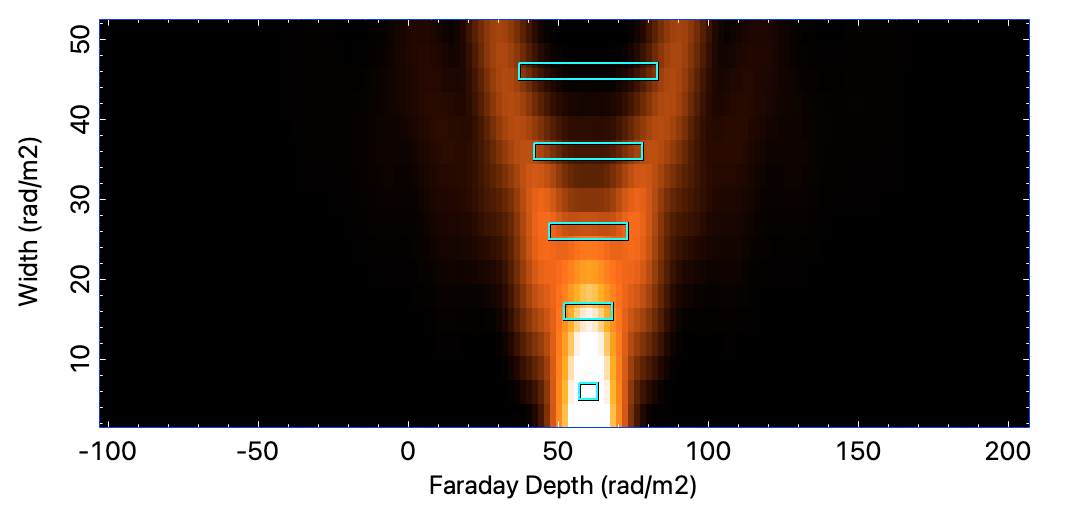} 
\caption{Faraday spectra along the horizontal axis at each width as indicated along the vertical axis.  The cyan boxes show selected \emph{input} tophat widths.  Top: \nomi; Bottom: \ful.  
\rev{The spectra for each width have been averaged
over 100 different realizations, each with a ramp of polarization angle ranges extending from
$0$ degrees  across the tophat to $90$ degrees across the tophat.}
\rev{At widths $\gtrsim 25~ $\radmm, the spectra are dominated by peaks at the edges of the input tophat, as opposed to its actual continuous distribution.}
} 
\label{TopHatWidth}
\end{figure}

A more comprehensive display of the  responses to the tophat continuous distribution is shown in Figure \ref{TopHatWidth}.  At each tophat width, we averaged 100 different tophat distributions with different gradients in polarization angle, ranging from a change in position angle across the distribution from 0 to 90 degrees.  The behavior seen in the 1D plots of Figure \ref{topslice} can also be seen here; a broadened Gaussian separating into two branches as the width increases. For \ful~ only,  the range of widths from  $\approx$10-25~\radmm~ ($0.6 - 1.5 \times \Phi_{full}$) shows suggestions of the input tophat distribution.

%An example of the effects of a variable intrinsic EVPA is shown in Figure
%\ref{TopHatRamp} for the widest (50 rad m$^{-2}$) tophat function.
%As the EVPA ramp steepens, the response over the bulk of the center of
%the tophat function reduces to a near zero level. causing the response
%to appear as two well resolved component.

\begin{figure}
    \centering
    \includegraphics[width=2.5in]{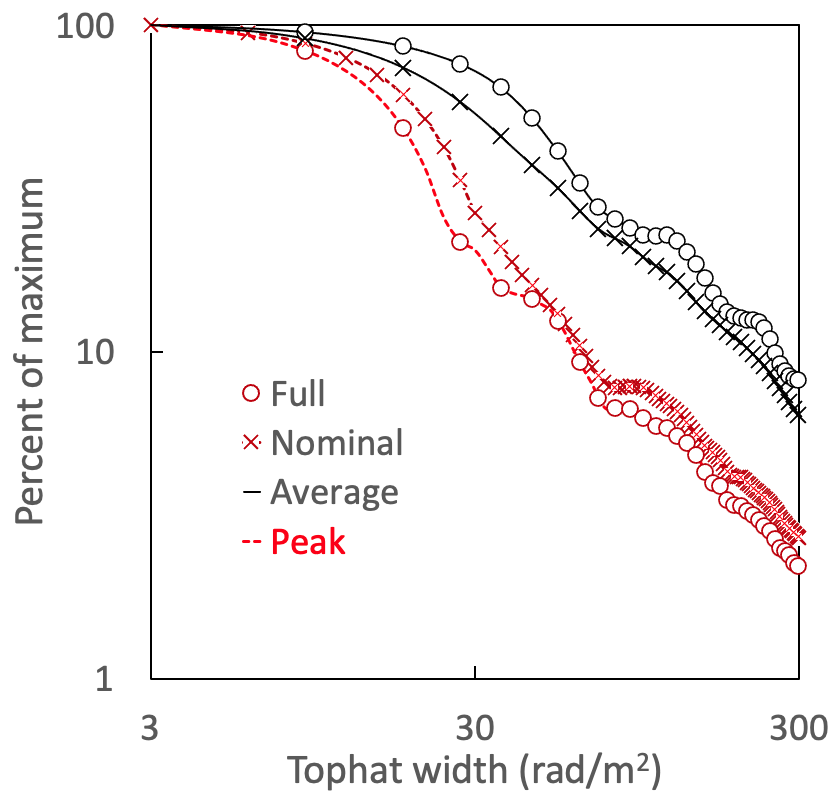}
    \caption{Recovered signal strength as a function of tophat width for \nomi~ and \ful, as described in the text.}
    \label{fig:maxscale}
\end{figure}

\rev{We now look quantitatively at the observed loss of power as a function of the width of the Faraday distribution.  We start by looking at the peak amplitude in the Faraday spectrum, and find that it falls by a factor of 2, for both \nomi~ and \ful, at $\approx 20$~\radmm~  (Figure \ref{fig:maxscale}}).  This is much less than the \emph{max-scale} of 109~\radmm~  predicted by \bdb.  We can even look at the average power over the entire spectrum, which would be difficult to do in practice, and find that the power drops by 2 at $\approx 40$~\radmm. Our new estimate of the appropriate width to half-power, $W_{max}$, is 25~\radmm, comparable to what is observed.  %As demonstrated by \bdb, the signal recovered in the Faraday spectrum drops as the width increases.  Here we compare the recovered signal strength measured in two different ways, the peak amplitude in the Faraday spectrum, and the average (or total) amplitude over the full spectrum.  The results for both \nomi~ and \ful~ are shown in Figure \ref{fig:maxscale}.  There is little difference between the two methods.  The calculated maximum scale, as described above is $\approx$109~\radmm.  However, we observe a much steeper falloff, reaching half power at $\approx40$~\radmm~ and $\approx20$~\radmm~ for the average and peak flux, respectively.  This is mostly or completely due to the inclusion of polarization angle gradients across the tophat distribution, which are included in our reported averages.
\begin{figure}
    \centering
    \includegraphics[width=1.6in]{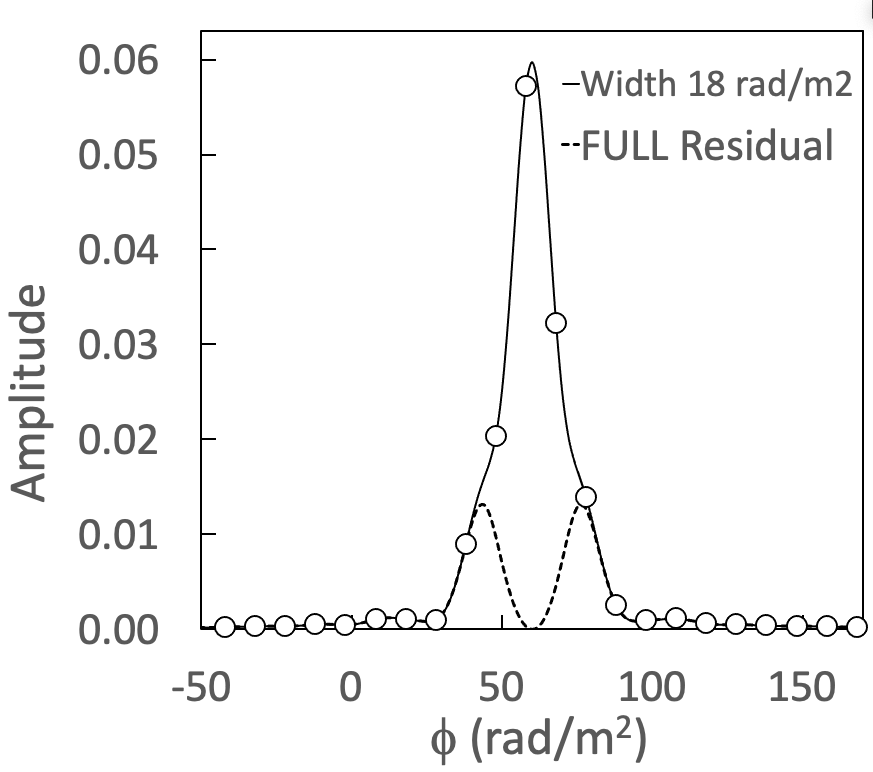}
  \includegraphics[width=1.6in]{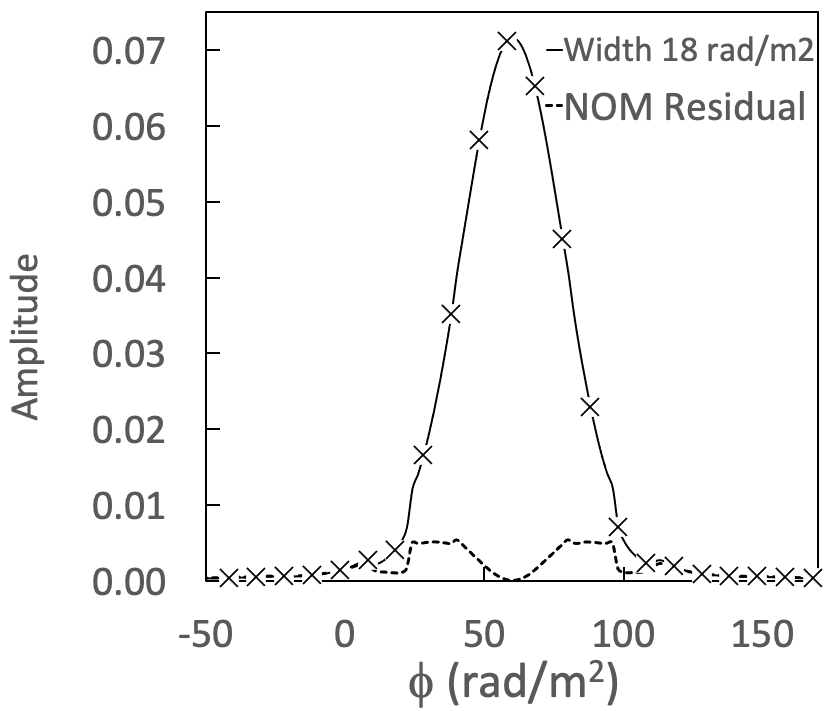}
    \caption{Faraday spectra for tophat input distribution showing original and residual spectra after subtracting out a single Faraday component with the observed peak amplitude. Left: \ful; Right: \nomi. }
    \label{fig:resid}
\end{figure}

We now address the critical question about whether \ful~ is superior to \nomi~ in identifying Faraday complexity, i.e., the presence of more than a single $\delta$ function component. We examine this by simply finding the peak in each spectrum and then subtracting it using the restoring beam of width $\Phi_{nom}$ or $\Phi_{full}$, as appropriate. \rev{This is equivalent to not restoring any clean components at and adjacent to the peak in the spectrum.} \footnote{\rev{An alternative process to search for complexity could use a loop gain of 1 in cleaning, and subtract out only one component;  we did not explore that option here.}}
%We note that the observed peak amplitude actually reflects the integrated flux over the the range of Faraday depths corresponding to the width of the Faraday beam;  however, the procedure of using the observed peak height and assuming it is due to a single component is how an observer would need to proceed.  
~ One example of this process, for an input width just slightly above $\Phi_{full}$ is shown in Figure \ref{fig:resid}.  
\begin{figure}
    \centering
    \includegraphics[width=2.8in]{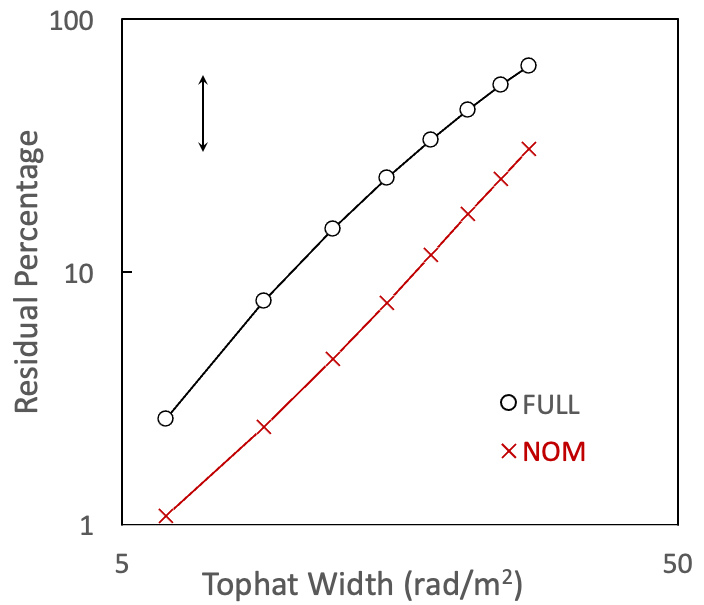}
    \caption{Percentage residual signal as a function of input tophat width,  integrated over Faraday spectrum after removal of single Faraday component, as described in the text, for both \nomi~ and \ful. The vertical bar indicates a factor of 2.}
    \label{fig:residratio}
\end{figure}
We then measure the total residual signal in the spectrum as a percentage of the total signal before subtraction.  As can be seen in Figure \ref{fig:resid}, the percentage residual is significantly smaller for \nomi~ than for \ful, \rev{as expected}.   This is unavoidable because the peak amplitude in \nomi~ represents an integration over a larger range of Faraday depths than for \ful.   In this particular \rev{example using the MeerKAT L-band, with} $\Phi_{nom}/\Phi_{full} \approx 3$, this leads to a \rev{factor of $\approx$2 improvement in} fractional residual power, i.e., the ability to detect complexity; \rev{this improvement applies for} input widths $\leq \Phi_{full}$, as shown in Figure \ref{fig:residratio} 

\subsubsection{\rev{Recovery of extended distributions}}
\rev{For continuous distributions of Faraday depth, the interference between components at different depths causes both the input power and the power recoverable in the Faraday spectrum to fall as a function of width.  The width at which the power falls by a factor of two is defined here as $W_{max}$, and its derivation is presented in the Appendix. To simultaneously have sufficient resolution to determine that the distribution has a finite width, and to have sufficient power to detect it, requires that  $\Phi < W_{max}$. Using Eqs. \ref{phinom}, \ref{wmax} and, again,  $\rho=\frac{\lambda_{max}}{\lambda_{min}}$, this requires for \nomi :}
\begin{equation}
 \rev{ 0.18~\frac{\rho^4-1}{\rho^2} > 1,}
\end{equation}
\rev{or }
\begin{equation}
\rev{  \rho > 6 .}
\end{equation}
\rev{The equivalent conditions for \ful, with Eq. \ref{phiful} are:}
\begin{equation}
   \rev{ 0.335~\frac{(\rho^2+1)^2}{\rho^2} > 1,}
\end{equation}
\rev{which is satisfied for all values of $\rho$. Thus, using \ful, continuous extended distributions in Faraday depth can always be at least
marginally resolved, e.g., as seen in Fig. 7.  However, among
the surveys listed in Table 1, using the \nomi~ resolution, only
SKA1-Low will be able to resolve detectable broad structures.}

The details of the shapes \rev{recovered with \ful} depend on what variations are present in the \rev{polarization angle across the Faraday distribution.}   In the simplest physical case of a spatially unresolved foreground patchy Faraday screen in front of a uniform polarization angle source, the polarization angles would be constant.  However, there can be arbitrary changes in polarization angle as a function of depth for mixtures of thermal and synchrotron emitting material.
%With a better understanding of the width of Faraday depths which are observable with a given frequency coverage. we can now ask about their resolvability, i.e., whether the full width of the Faraday depth distribution where the power has dropped by half larger than the Faraday resolution, $\Phi$, as given in the main text.  We define the quantity \emph{resolving power, RP} , where $RP_{nom,full} = 2.3\sigma_{0.5}/\Phi_{nom,full}$.  This result depends only on $\rho$, and is given by
%\begin{equation}
%RP_{nom} = 0.175(\rho^2-1)(\rho^{-2}+1),~ ~  for ~ \Phi_{nom}, 
%\end{equation}
%\begin{equation}
%RP_{full} = 0.33(\rho^2+1)(\rho^{-2}+1),~ ~ for~  \Phi_{full}.
%\end{equation}

%Ideally, one would want RP to be at least a factor of a few, to get a sense of the shape of the Faraday depth distribution.  $RP_{nom} > 1$ only for $\rho~>~2.43$, and this requirement is usually not satisfied, as shown in Table \ref{surveys}.  $RP_{full}$, by contrast is always $>~1$.   Thus, if information about the shape is required, then one must use \ful. 

\subsection{Two Faraday Components}
{\em {Key Findings}: Two Faraday components can often be distinguished at separations less than $\Phi_{nom}$, \rev{with improved detectability using \ful.}  In this regime, the separations, amplitudes and polarization angles are, however, not accurately recovered.  At some separations, spurious emission at the \rev{mean} depth reappears, but is much weaker in \ful~ than in \nomi.} %Detectability of emission from a second component is also enhanced in \ful.}
 
\begin{figure}
    \centering
    \includegraphics[width=3.2in]{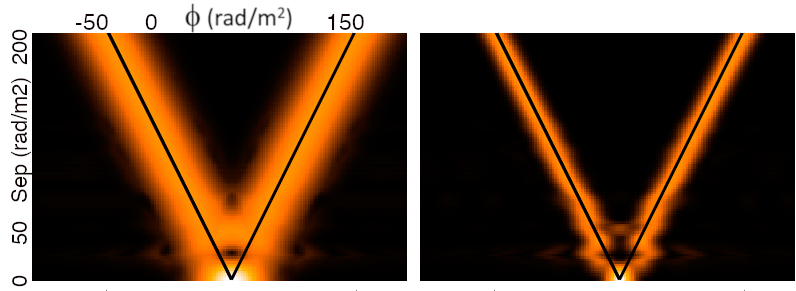}
    \caption{Faraday spectra along the horizontal axis, \rev{with increasing separations between two components with the same polarization angle at each higher row,};  this is similar to the display in Figure \ref{TopHatWidth}.  The left column shows the results from \nomi, and the right column from \ful. \rev{Note the spurious power that sometimes appears at the mean depth.}\revv{Such spurious components are the result of the interference that occurs in the presence of phase variations with Faraday depth, such as noted in Figure 3.}}
    \label{sum2}
\end{figure}

\vspace{0.15in}
We now turn to the second simple case,  two separated Faraday components with equal amplitudes.  We vary both the separation in depth between the components and the difference in their polarization angles.  Again, since the science often requires us to maximize the amount of Faraday structure we can see, we test separations that are both smaller and larger than $\Phi_{nom}$.  A global view of the results in shown in Figure \ref{sum2}.
\begin{figure}
\centering
\includegraphics[width=2.8in]{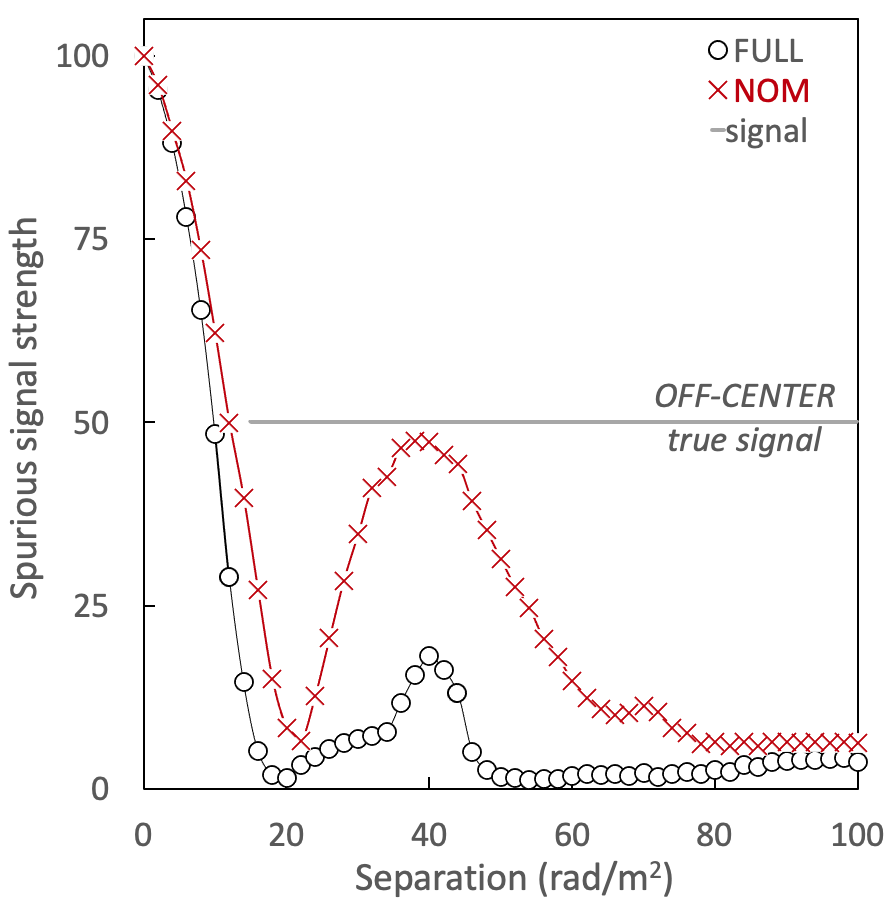}
\caption{``Spurious" power \rev{for \nomi~ and \ful,} observed at the mean Faraday depth as a function of the separation in depth between two components with amplitudes=50; \rev{the horizontal line shows the true signal strength that should be observed away from the central peak.}}
\label{spurious}
\end{figure}

At the bottom, where the depth separations are 0, the spectrum peaks at the mean Faraday depth of 60~\radmm, as expected.  At large separations, \rev{at the top of the figure}, the two individual components are easily visible, with respective depths that track the input depths, as expected.  

The behavior in between these two extremes is quite complex. We first look at the observed power at the mean depth, which should start at the sum of the amplitudes of the two components and then fall to zero as they are well separated.  If the combined Faraday spectrum were simply the sum of the two input \emph{amplitude} spectra (which it's not), then this falloff would follow a Gaussian shape.  Instead, since it is the complex spectra that are combined, there is interference between the two components which depends on their relative phase.  \rev{This power at the mean depth was studied by \cite{Kuma14}, who called it a ``false signal"; they showed that its strength was a function of the separation, relative phase and relative amplitudes of the two components, similar to the findings from these studies.}

\rev{The additional information shown here is that the spurious power at the mean depth also depends on the restoring beam.}  After averaging over all relative phases \rev{between the two components}, we show the observed \rev{spurious} power at the mean Faraday depth in Figure \ref{spurious}. \rev{Our results are similar to those of \cite{Kuma14}, although we average all the power in the restoring beam centered at the mean depth, while they select only limited clean components.} There is a region around $\sim40~$\radmm~ where the two components interfere to produce power at the mean depth; this spurious power is much stronger  in \nomi~ than in \ful, \rev{and also extends over a much larger range in depth separation.}\revv{Such interference arises in the presence of phase gradients as a function of Faraday depth, which can even arise in the cleaning process, such as noted for Figure 3.}  %The bottom three rows of Figure \ref{sum2} shows that the depth at which this degeneracy occurs is dependent on the relative phase between the two components.

\begin{figure}
    \centering
    \includegraphics[width=2.5in]{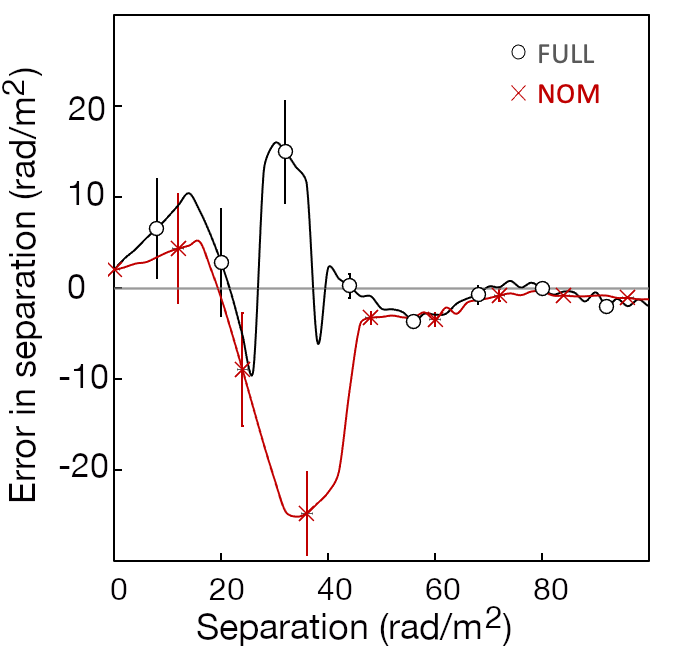}
%    \vspace{0.01in}
%    \centering
%    \includegraphics[width=2.5in]{figslr/TwoNoNoise0sepNOV.png}
    \caption{Deviation from the observed separation in Faraday depths of the peaks in the spectrum as a function of the input separation.  The black \revv{and red} lines show the mean, averaged over all relative polarization angles between the two components, and the \rev{error bars indicate} the rms scatter.}
    \label{twosep}
\end{figure}

In addition to the  \rev{presence of }spurious signals, there are also deviations in the observed parameters \rev{from the input parameters for separations up to scales of $\approx \Phi_{nom}$}.   This  can be seen most clearly in the \rev{ non-monotonically increasing separations of the two components of the} \ful~ spectra of Figure \ref{sum2}; similar behavior is present in \nomi~ \rev{although it is less obvious}.  These deviations \rev{from the input separations} are also shown quantitatively in Figure \ref{twosep}, where the problems at separations less than  $\Phi_{nom}$ are clear.   Similarly, the rms deviations in observed amplitude (\rev{and} position angle) are  much higher at small separations --  $\sim33\%\ $  (and~ $20^{\circ}$) \rev{for separations $<\Phi_{nom}$, dropping to } $\sim10\%\ $ ( and~ $3^{\circ})$ \rev{for separations $>\Phi_{nom}$}.  The results are similar for both \nomi~ and \ful. 

The bottom line from all this is that two components with separations less than $\Phi_{nom}$ are detectable some of the time, \rev{depending on their relative polarization angles} but their observed parameters are not trustworthy.  Above $\Phi_{nom}$, \nomi~ and \ful~ perform equally well. \rev{At separations comparable to $\Phi_{nom}$,~ } \nomi~ \rev{shows considerably stronger spurious signals.}
\begin{figure}
    \centering
    \includegraphics[width=3in]{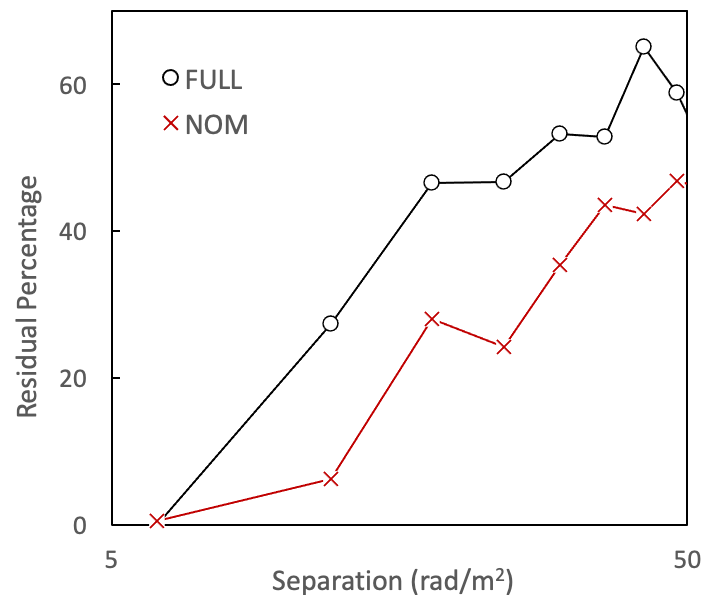}
    \caption{Residual power remaining in the clean spectra, as a function of component separation, after removal of a single component, expressed as percentage of the original power.}
    \label{residtwo}
\end{figure}

\subsubsection{Detectability of multiple Faraday components}

We can \rev{again ask the simpler question most relevant to the analysis that would be performed in surveys, i.e. what is} the detectability of Faraday structure as the separation between two components increases?  We again subtract out a single component from the \rev{peak in the} clean spectrum and measure the percentage of power remaining from $-\Phi_{full}$ to $+\Phi_{full}$ (i.e., two beam-widths), and the same for \nomi, using $-\Phi_{nom}$ to $+\Phi_{nom}$. The results are shown in Figure \ref{residtwo}.  At very small separations, the two components overlap and the expected residual is 0\%, as observed \rev{for both restoring beams}.   

At sufficiently large separations, we expect 50\% of the power to remain after subtraction of a single component, as observed.  Between these two extremes. \ful~ shows a higher percentage of residual power, up to a factor of $\sim$2, thereby increasing the detectable range for additional Faraday structure. \rev{This result is the natural, and perhaps obvious,  consequence of using a smaller restoring beam;  nonetheless, this quantifies the advantages to that approach for separations from $\approx 0.5 - 1.5 \times \Phi_{nom}.$} 

\section{Faraday Mapping}
In the simplest case,  only single Faraday components are present at each position in an image, and the spatial variations in Faraday depth $\Delta\phi$ are small with respect to the resolution $\Phi$ over scales comparable to the angular resolution (the ``beam"). In this case,  all of the information is present in maps of the peak Faraday depth in the spectrum for each pixel. Even in this idealized case, however, a series of two-dimensional maps at each Faraday depth (tomography images), are often useful to detect the spatial patterns.   Two-dimensional images in Faraday depth vs. position space ($\phi$ vs. $x$), where the orthogonal position has been fixed, provide another useful diagnostic.  Both of these techniques are exploited, e.g, in the recovery of the 3D structure of 3C40B using MeerKAT observations \citep{Rudnick2022}.  

These mapping techniques become essential when the variations $\Delta\phi$  are comparable to $\Phi$ over scales of the beam; in this case,  there is no longer a single Faraday depth at each position, and the Faraday spectra will be subject to the distortions examined earlier.  In this section, we use simulations to compare how Faraday tomography mapping and depth vs. position mapping appear in \nomi~ and \ful, and look at real Faraday structure data from MeerKAT.

\subsection{Simple Faraday Depth gradients}\label{gradcartoon}
{\em {Key findings:} Faraday tomography maps are {\bf spatially} broadened by the effects of finite Faraday resolution in the case of spatial gradients in Faraday depth.}

%In addition, both the finite size of $\Phi$ and the spectral distortions will translate into the Faraday tomography images, i.e., {\bf variations in $\phi$ translate into broadening the tomography images in the plane of the sky and in the $\phi$ vs. $x$ plane.}   In this section, we will examine these effects using simulated signals and MeerKAT observations from two clusters.

\begin{figure}
\centering
\includegraphics[width=3.5in]{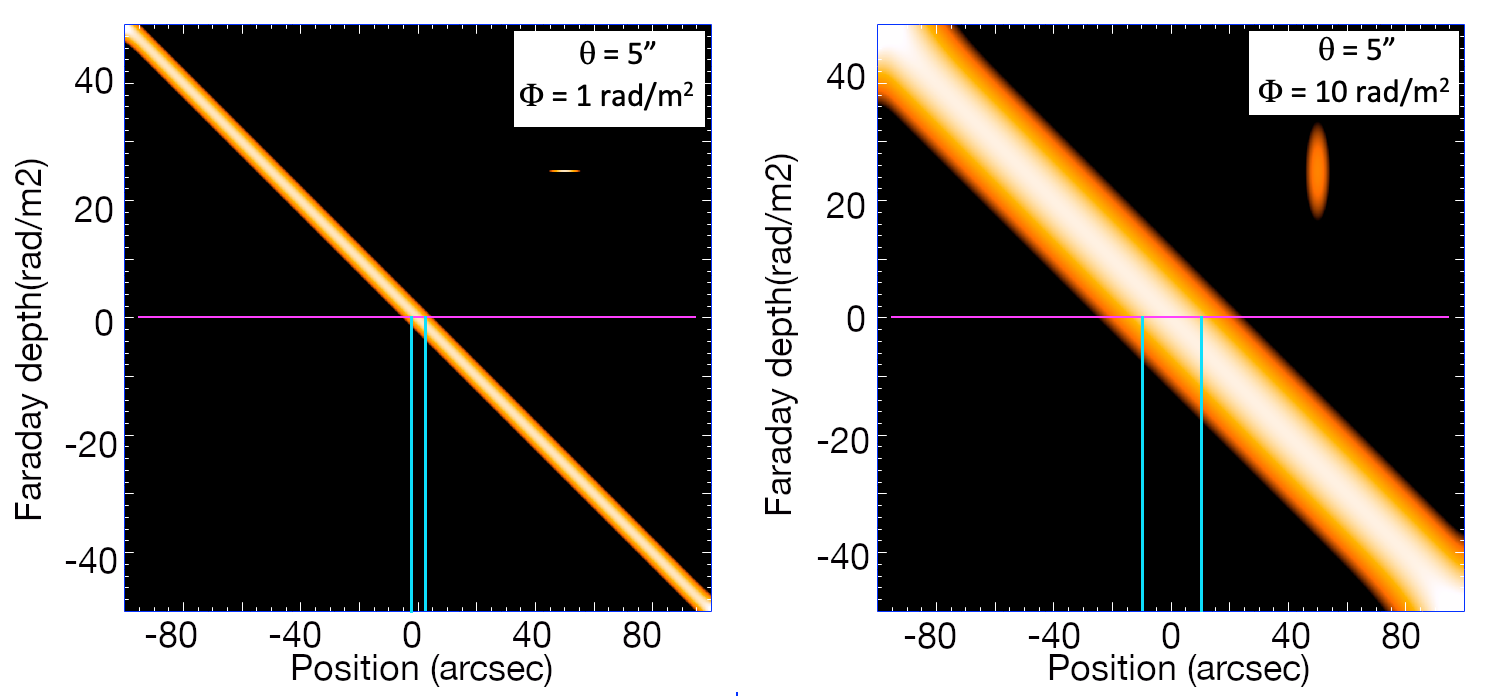}
\caption{Cartoon showing $\delta$-function Faraday spectra with depth as a function of position on the sky. In both panels, the spatial resolution $\theta = 5\arcsec$. Left: Faraday resolution $\Phi=1$~\radmm; Right: $\Phi=10$~\radmm.  The vertical lines show the observed \em{spatial} extent of the emission in the $\phi=0$ Faraday tomography (1-D) image. \rev{The 2D beam is shown in the upper right portion of each panel.} }
\label{faradayexamp}
\end{figure}

We start with a simple cartoon to illustrate how, in the presence of Faraday depth variations,  broadening of the Faraday spectrum leads to broadening in the plane of the sky. Figure \ref{faradayexamp} presents the amplitude of $\mathcal{F}(\phi)$ as a function of a single position coordinate. The Faraday spectrum is a simple delta function, whose depth changes linearly with position. The small bar in the upper right shows the "beam" in this depth vs. position space. One can examine the spatial distribution at a given Faraday depth by taking a 1-D slice at that depth as a function of position (the horizontal magenta line in Fig. \ref{faradayexamp} -- this is the equivalent of a tomography plane in 2-D). We then measure the spatial width between the half-power points (the vertical cyan lines). On  the left, the width of the feature at $\phi=0$~\radmm~ is 5\arcsec, as set by the spatial beam size.  On the right, the Faraday spectrum has been broadened to 10~\radmm. Despite the spatial beam still remaining at 5\arcsec, as can be seen in the beam in the upper right \revv{of the frame}, the \emph{observed} width at $\phi=0$~\radmm~ is now  \rev{$\approx$}20\arcsec.   This spatial broadening occurs because the tomography cut (plane) at a depth of 0~\radmm~ is actually measuring the emissions from $\pm~5$~\radmm, which come from different positions.  This broadening therefore blends and masks features in tomography images, compromising the hard-won spatial resolution set by the telescope. 

\rev{Since the Faraday beam broadening is convolved with the original spatial beam, the effective size of the spatial beam in a tomography image, designated here as $\theta_t$, can be approximated as}
\begin{equation} \label{thetat}
\rev{\theta_t \approx \sqrt{ \theta_0^2 + (\frac{dx}{d\phi}\Phi)^2} }
\end{equation}
\rev{where $\theta_0$ is the spatial beam size, $x$ is the position coordinate,  $\frac{dx}{d\phi}$ is the local spatial gradient in Faraday depth and $\Phi$ is the Faraday restoring beamwidth.  In the above case, $\frac{dx}{d\phi}$ is $2\frac{\arcsec}{rad m^{-2}}$ 
and $\Phi = 10 ~$ \radmm ,
so~ $\theta_t\approx 20.6\arcsec$, as observed. }
If the Faraday depth variations are not a simple gradient, then the spatial effects due to Faraday depth variations may not be easily recognizable, as we will see with actual data, below.
 
 \subsection{Faraday Variation Grid}
{\em{Key findings:} In Faraday tomography mapping, spurious spatial structures appear when multiple Faraday depths are present within the \revv{spatial} beam.   \revv{Use of the narrower restoring beam $\Phi_{full}$} allows  a larger range of Faraday widths to be free of such spurious structures. }

\vspace{0.15in}
\begin{figure}
    \centering
    \includegraphics[width=3.5in]{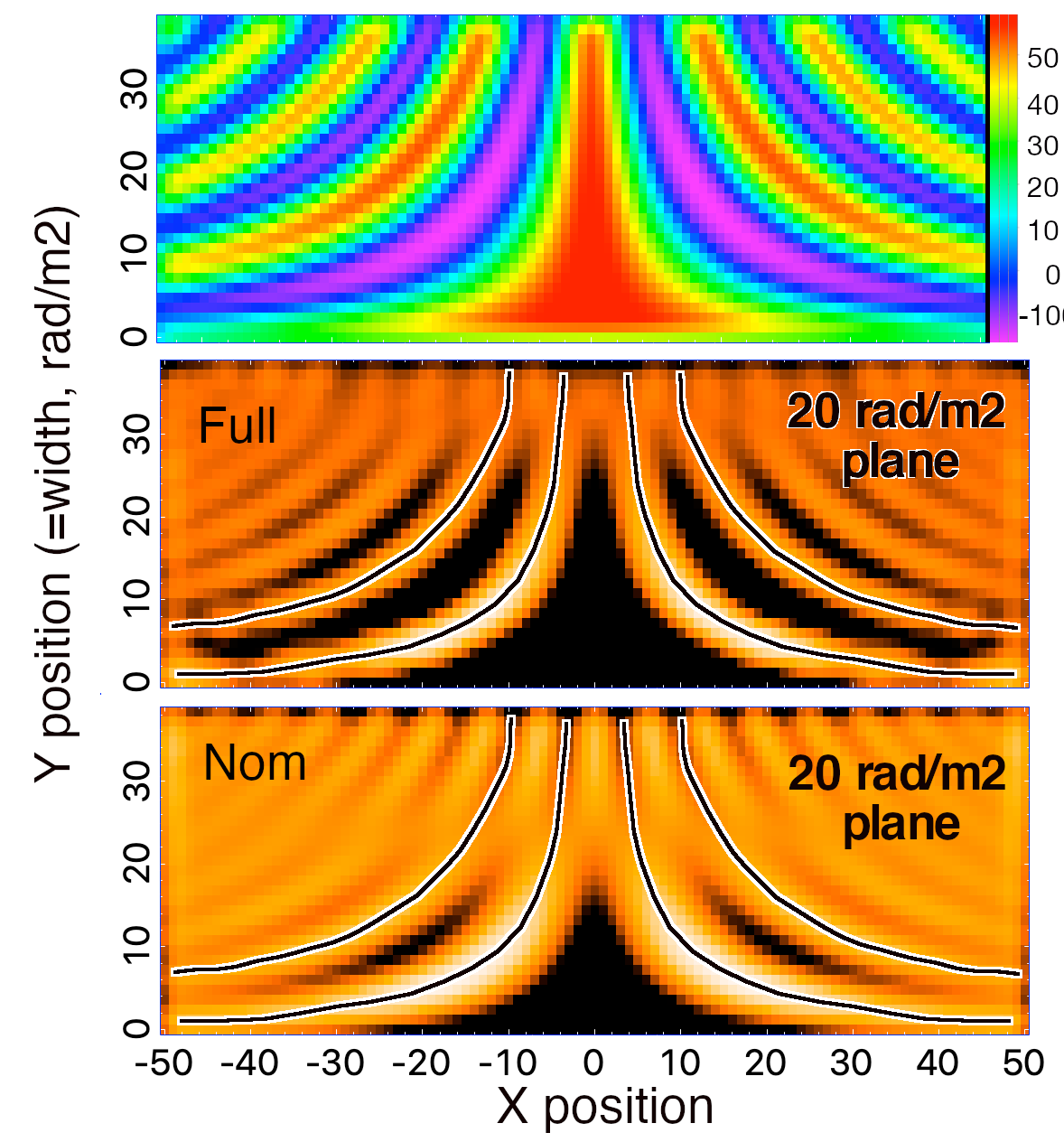}
    \caption{Top: Faraday depth in plane of sky.  Middle: Single tomography plane at depth of 20~\radmm~ from \ful, intensity in heat, each row normalized to the same average.  Bottom: Same for \nomi.  The X and Y coordinates are positions in the sky; the magnitude of Faraday depth variations changes with Y, as described in the text. The black lines with white borders shows the location of input depths of 20~\radmm, for the first two sinusoidal patterns}
    \label{228}
\end{figure}

The above cartoon, Figure \ref{faradayexamp}, illustrates the overall broadening effect, but, for simplicity,  assumes that the broadening took place in the Faraday {\em amplitude} spectrum.  More accurately, the broadening occurs in the complex Faraday space, and so the resulting patterns are more complicated.  As a simple illustration of what happens in the map plane when there is mixing of different Faraday components, we create  cubes in (x, y, frequency) space, where every pixel has a single Faraday depth. The top panel in  Figure \ref{228} shows the Faraday depth at each position.  Along each row, the Faraday depth changes sinusoidally between the values of -20 and +40~\radmm.  The spatial wavelength increases from the top row to the bottom row so that there are $\sim8$ full cycles on the top row, and zero cycles on the bottom row.  We then smoothed the values of Q and U along the X-axis by a Gaussian with 3 pixel FWHM.  Each smoothed pixel thus has contributions from a range of Faraday depths.  Along the bottom row all of the pixels have a depth of 20~\radmm, so the smoothed Q and U have only that single Faraday depth.  Along the top row, the range of Faraday depths in each pixel varies from 6~\radmm~ to 36~\radmm.  The maximum Faraday width in each row is given along the Y-axis.
 
 This simulated Faraday grid is equivalent to having $\sim$3 dominant Faraday components in each pixel.  Its effects differ in detail from the idealized continuous distributions and two-component cases discussed above.  Nonetheless, it gives a simple overview of how \nomi~ and \ful~ behave in Faraday tomography images with variations in the amount of Faraday structure.
 
 The middle and bottom panels show  a single tomography image plane at the Faraday depth of 20~\radmm, for \nomi~ and \ful, respectively,  cleaned and restored as with our earlier experiments. The black lines with white borders show the the locations of input Faraday depth of  20~\radmm, for the middle cycle of the sine waves.  In the idea world, the tomography images at 20~\radmm would simply trace the black/white lines; they do not.
 
 The first finding is the observed bands in the tomography images are much broader than the 3-pixel smoothing beam, the same width as the black/white line.  That broadening, \revv{present in all rows},  is because the 20~\radmm~  image actually samples emission from $\sim$12~-~ 28~\radmm~ ($\sim$-2~-~42~\radmm~) for \ful~ (\nomi), respectively.  \revv{In the row where} the Faraday mixing width is $\sim$10~\radmm, the \revv{broadening causes the} band in the \nomi~(\ful) image \revv{to be} 16 (8) pixels wide, respectively, instead of the spatial beamwidth of 3 pixels. \rev{This is another example of the broadening described by Equation \ref{thetat}.} 
 
 Another consequence of the broadening  is that the ratio between the brightest and the faintest features are reduced in the tomography image.  \rev{This is because the broadening is a function of position on the image, since the local Faraday depth gradients vary.}  At a Faraday mixing width of 10~\radmm~, the ratio of brightest/faintest is 5.2 (575) for \nomi~ (\ful), respectively.  For mapping purposes, \ful~ is clearly superior.

  In some ways, these broadening effects are similar to what happens in  total intensity images.  Based on our sampling of the intensity restoring beam, adjacent pixels are not independent, but represent the emission at the exact location of that pixel as well as emission occurring at adjacent pixels. \rev{This is universally understood in the radio astronomy community, so there is no confusion. However,  in tomography images, } what appears in any pixel in a tomography plane also reflects the emission in the planes between $\pm\Phi/2$, where $\Phi$ is the Faraday beam. If one were examining all the planes together, as in a cube or a movie display, the origins of the broad structures would be apparent.  In a single tomography plane, there is no way to distinguish between intrinsically broad spatial structures and observed spatial extent due to the Faraday broadening. 
 
 The situation in this Faraday/spatial broadening is actually more complicated, however, by the fact that the blending takes place in the vector space of Q and U, so that both constructive and destructive interference can appear.  This results in spurious structures in the Faraday spectra, as described in the previous two sections, mapping into spurious structures in the Faraday tomography image planes.

The spurious structures are apparent by noting  that the observed bands in the 20~\radmm~ tomography images track the 20~\radmm~ input locations, but only when the mixing widths are sufficiently small.  As the widths increase, increasing amounts of power are found between the locations of the 20~\radmm~ inputs, until finally all the power is at the spurious locations of the 0 and 40~\radmm~ inputs.  Equal or greater power at the spurious locations is found for widths above 20~\radmm~ (30~\radmm) for \nomi~(\ful), respectively.  Thus, there is a considerably larger range of Faraday widths where \ful~ is free of spurious structures.
 
  Thus, in real maps, when significant Faraday variations occur within a spatial beam, spurious structures can appear in the tomography images. Since Faraday beam depolarization is almost ubiquitous, the potential for spurious structures in tomography images is very high. These spurious structures do not correspond to any real structure at the respective Faraday depth.  In addition, as seen earlier, the breadth of the Faraday beam will also cause structure to appear at Faraday depths where they are not present. While this is true for \ful~as well, it is several times worse for \nomi.

\subsection{MeerKAT polarization mapping}
{\em {Key findings}: In the presence of complicated Faraday structure in both depth and the plane of the sky, images of the peak depth of the Faraday spectrum are almost identical using \nomi~ and \ful. However, the broader \nomi~beam  blurs out detailed Faraday variations that are seen clearly in \ful. In addition,  \nomi~ produces spatial structures in  Faraday tomography planes where \ful~ shows there is no emission.  }
\vspace{0.15in}
\begin{figure}
    \centering
    \includegraphics[width=3.5in]{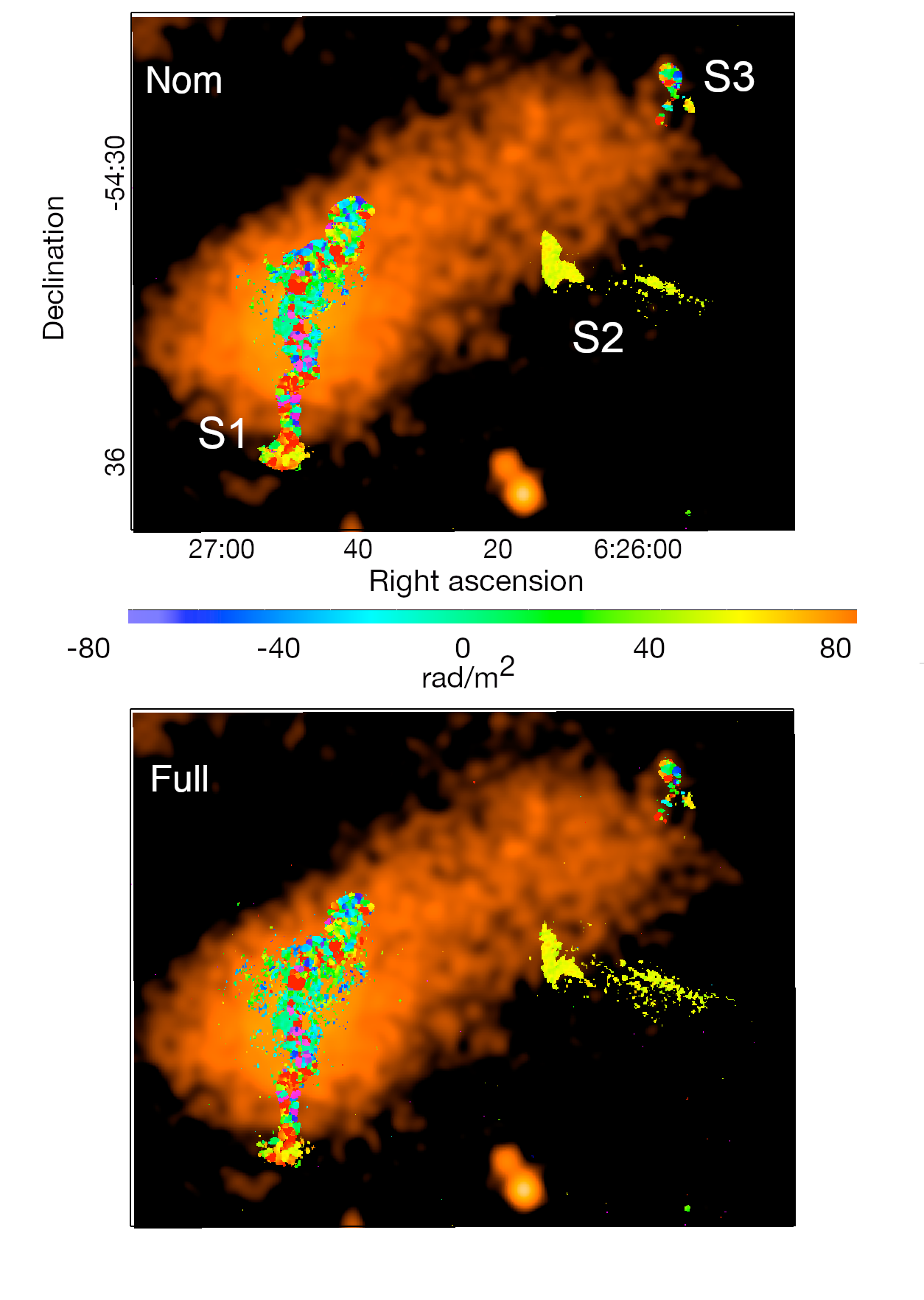}
    \caption{Faraday depth at the peak amplitude in the northern section of Abell~3395, showing all pixels with brightness $>35~\mu$Jy/beam. The results are for \nomi~(top) and \ful~(bottom).  The underlying heat image shows the X-ray brightness from eROSITA \citep{3395a,3395b}, with this version courtesy of Angie Veronica.}
    \label{sel3395}
\end{figure}

MeerKAT observations of the cluster of galaxies J0627.2-5428 (Abell
3395) were reported by \cite{Knowles2021}.  More detailed analysis of this field, and comparison with X-ray data from eROSITA, were presented by \cite{3395a} and \cite{3395b}.  The observations consisted of
approximately 9 hours duration, including calibration, at L band
(856-1712 MHz).
Calibration was described in \cite{Knowles2021}.
The data were imaged in Obit/MFImage with 0.3\% fractional bandwidth (123
spectral channels) using joint Q/U deconvolution.  The beam size was 6.8$\times$6.73\arcsec at an angle of 88$^o$, and the pixel size was 1.194\arcsec. 
A Faraday spectrum cube was generated using RMSyn with 2~\radmm~  sampling between -600 and +600 rad m$^{-2}$ and
deconvolved/restored as described above for both \nomi~and \ful. Following the standard analysis, we created 2D images by finding the depth and amplitude of the peak in the Faraday amplitude spectrum for each pixel.  The rms scatter in the peak amplitude image was $\sim$2.5$\mu$Jy for both \nomi~ and \ful.

Figure \ref{sel3395} shows the Faraday depths at the peak in the Faraday spectrum ($\phi_{peak}$) for each pixel, wherever the signal:noise was $>14$.  The results for \nomi~ and \ful~ were virtually identical, as seen in this figure and in Table \ref{scatter}; the small differences come largely from the handful of isolated pixels due to enhanced noise around the bright sources. Sources S1 and S3 show much larger scatters in $\phi_{peak}$ than in S2.  This is consistent with S1 and S3 being embedded or behind the broad band of X-ray emission that connects them in projection; there is, however, no independent confirmation available.  On larger scales, the mean  RM  in this region \footnote{From the CIRADA RM cutout server, http://cutouts.cirada.ca/rmcutout/} is influenced by cluster sources themselves, so the Galactic foreground is not well constrained. 

\begin{figure}
\centering
\includegraphics[width=3.5in]{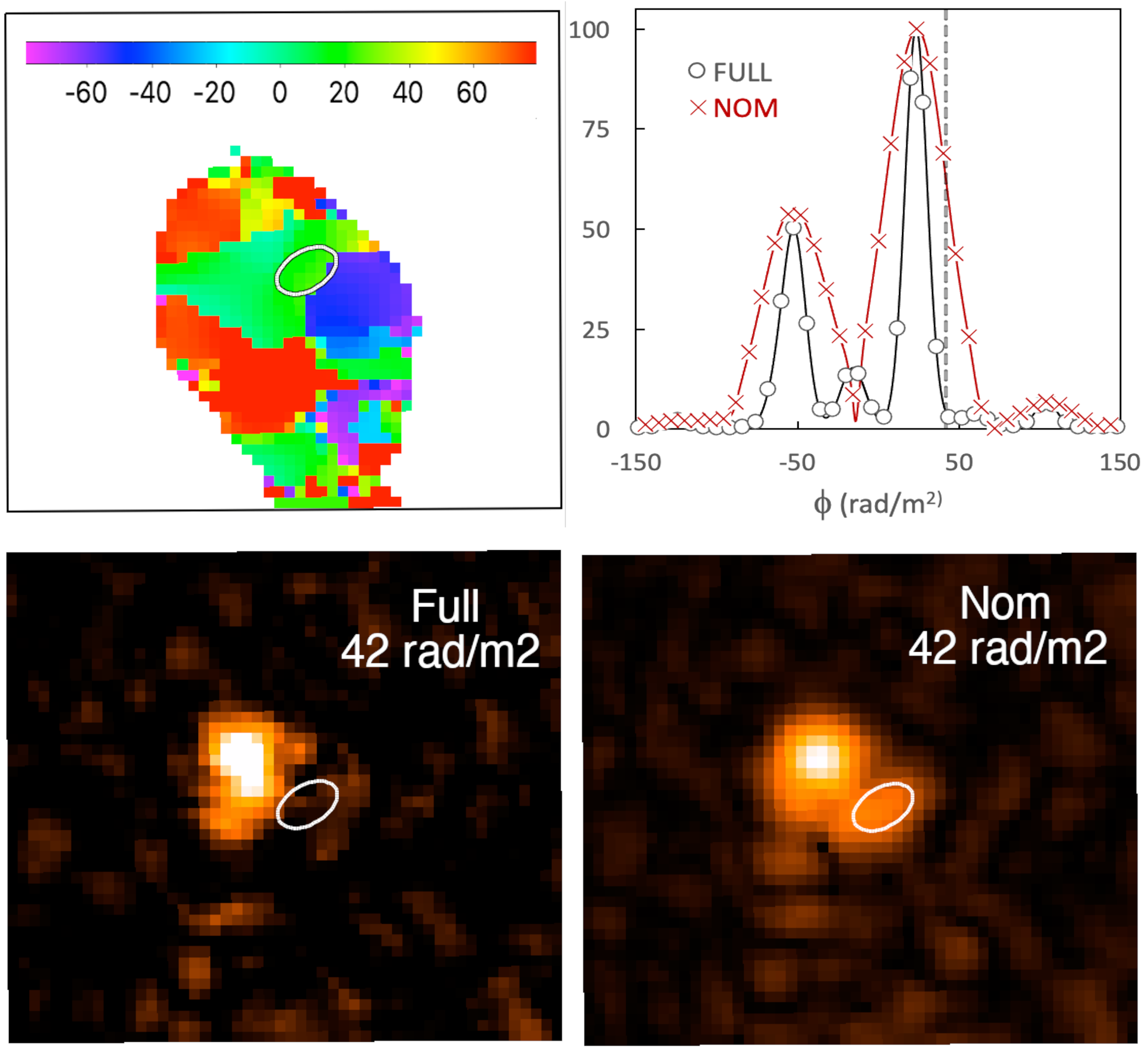}
%\caption{Faraday structure around the top bright portion of S3. The box is 62\arcsec x 56\arcsec centered at 06 25 56.1, -54 27 54.  Top left: $\phi_{peak}$, with the white ellipse showing region of interest; Top right: Faraday spectra from the ellipse using the same symbols as before.  Dotted vertical line indicates $\phi$=42~\radmm~.  Bottom:  Faraday tomography images at $\phi$=42~\radmm~ from \ful~(left) and \nomi~{right).}
\caption{North end of S3.  Top left: Peak Faraday depth; the depth at the ellipse is 21~\radmm;  Top right: \nomi~ and \ful~ spectra at the position of the ellipse. Vertical line indicates depth of 42~\radmm.  Bottom left: \ful~ tomography plane at 42~\radmm. No emission is seen in the ellipse, as expected from spectrum. Bottom right: \nomi~ tomography plane at 42~\radmm. Emission seen at ellipse from emission peaking at 21~\radmm. \rev{The peak visible in the \ful~ spectrum at -15~\radmm~ comes from a strong component at that depth just to the east of the ellipse and not visible here;  a small amount extends into the ellipse for \ful, although not for \nomi, reflecting the slightly different interference in the two cases.}}
\label{S3}
\end{figure}

 In Figure \ref{S3} we show one example of how a broader beam can create ``spurious" structures in Faraday tomography images, similar to some of those seen in the simulations discussed earlier. Structure at the ellipse appears in the \nomi~ \rev{42 \radmm~ tomography} image, but not in the \ful~ image.  The spectrum at this location, in the top right of the figure, shows why.  At this location, there is a bright peak in the Faraday spectrum at $\phi$=21~\radmm~. At $\phi$=42~\radmm~ there is still power in \nomi~ from the $\phi$=21~\radmm~ peak, so the tomography map shows a bright patch there. However, in \ful, the emission has dropped to near 0 by $\phi=$42~\radmm, so there is no feature at the ellipse. 
 \begin{table}
\vskip 0.01in
\caption{Peak Faraday depths in Abell 3395}
\vskip 0.01in
\begin{center}
\begin{tabular}{|c|c|c|c|c|}   
\hline
\hline
\multicolumn{1}{|c|}{Source} & \multicolumn{2}{c|}{\nomi} & \multicolumn{2}{c|}{\ful}\\
\hline
 & $\langle\phi_{pk}\rangle$ & $\sigma_{\phi}$  & $\langle\phi_{pk}\rangle$ & $\sigma_{\phi}$ \\
           & \radmm &  \radmm & \radmm & \radmm \\
\hline
S1 & 25 & 84 & 29 & 88\\
S2 & 56 & 9 & 50 & 4\\
S3 & 43 & 57 & 39 & 55\\
\hline
\end{tabular}
\end{center}
\hfill\break
\label{scatter}
\end{table}
%The larger value of $\Phi_{nom}$, however, means that there is still high signal:noise emission here at $\phi$=42~\radmm~. 

If one were viewing the full Faraday cube, it would be obvious that this 42 \radmm~ emission comes from a different depth.  One could avoid this problem by only sampling the Faraday tomography images 3$\times$ more sparsely.  However, and this is the key issue, this comes at the expense of losing information about the spatial/spectral Faraday structure that is present in the cube.  \rev{The use of smaller restoring beams, as in \ful, does not remove the problem of the smearing of structures from one depth map to another, it simply reduces the range of depths over which this is a problem.  In all cases, claims of emission at a specific Faraday depth require examination of the full cube.}
%\begin{figure}
%\centering
%\includegraphics[width=3.5in]{figslr/3C40S_RMfullnom.png}
%\caption{Faraday structure in the southern lobe of 3C40B, constructed from Faraday tomography images at 2~\radmm~ (red), 10~\radmm~(green), and 18~\radmm~(blue).}
%\label{3C40S}
%\end{figure}
\begin{figure}
    \centering
    \includegraphics[width=3.5in]{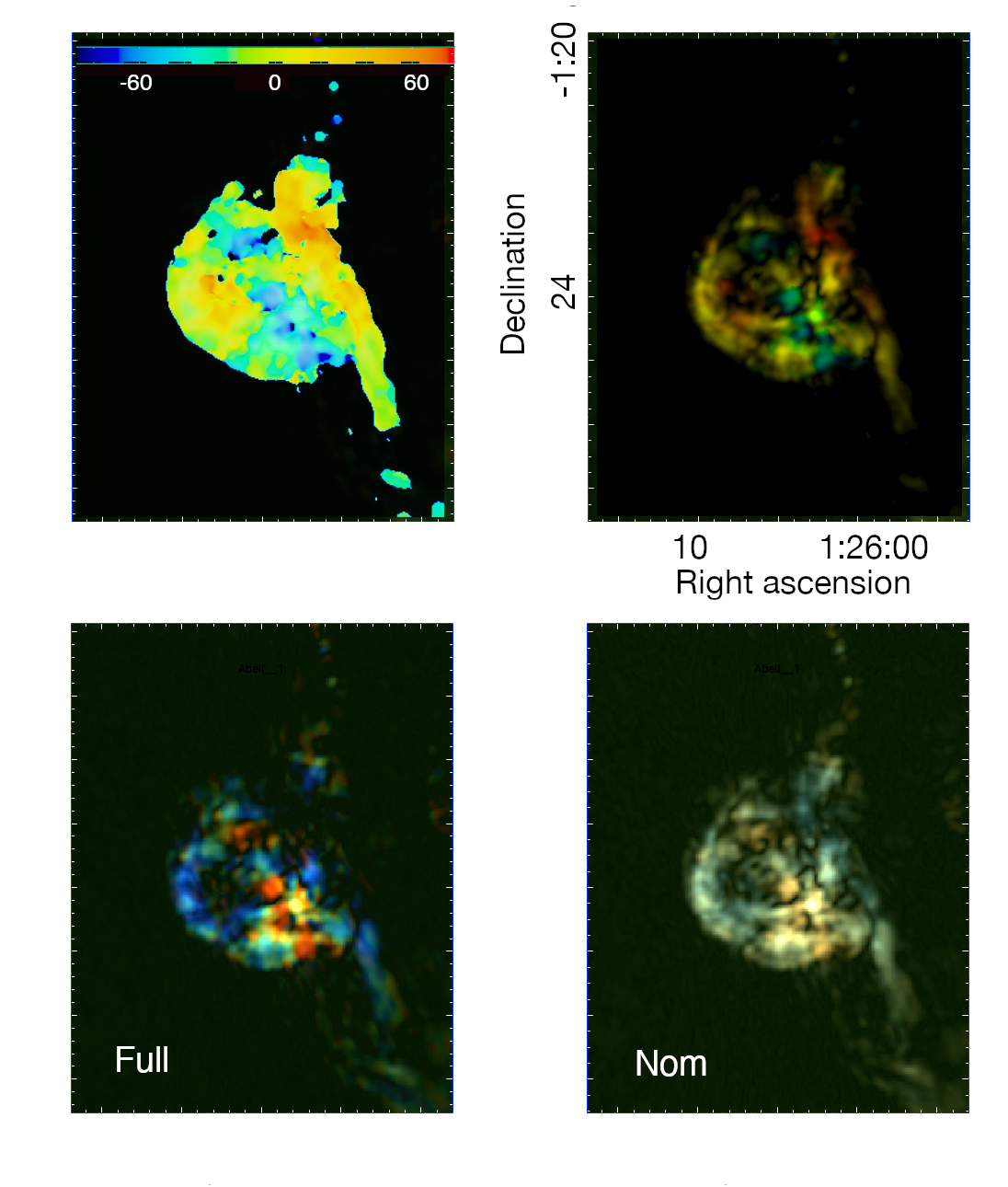}
    \caption{Faraday structure in the southern lobe of 3C40B. Top left: Peak Faraday depth. Top right: Peak depth color coded, as in top left panel, brightness corresponding to amplitude at peak depth. Bottom left: Partial structure as seen in  Faraday tomography images at 2 (10, 18)~\radmm~  in red (green, blue), for \ful. Bottom right: same as left, but smoothed to \nomi~resolution.}
    \label{3C40S}
\end{figure}
 An example of lost information using \nomi~ can be seen in the southern lobe of 3C40B, which is analysed in \cite{Rudnick2022}, \revv{using similar procedures to Abell~3395, as discussed above.}.  There, using \ful, movies of the Faraday cube show that the lobe is comprised of long thin coherent structures at different Faraday depths, which likely indicate different distances along the line of sight, allowing us to interpret the 3D structure.  
 
 Several views of this lobe are shown in Figure \ref{3C40S}.  On the top left is the image of the peak depth at each pixel;  this is the standard way depth (RM) maps are shown, and it obscures all of the detail in the lobe. \rev{Such a display is most useful when the variations in Faraday depth are dominated by a foreground screen, and the details of the lobe are irrelevant.} On the top right is the same image of peak depth in color, where the brightness shows the intensity of the amplitude at the peak depth (polarized intensity).  The various structures are immediately visible, \rev{because in this case, the depth is connected to the lobe structures themselves, as can be clearly seen in the movies in \cite{Rudnick2022}. It is important to realize that this type of display could be misleading if, in fact, the depth variations \revv{were} due to a foreground screen.  Examination of the full Faraday cubes is essential to separate foreground from local effects .}  Three \ful~ tomography planes, from the low Faraday depth part of the distribution are shown in the bottom left. % Extended features at the same Faraday depth are apparent. Looking more closely at the NW edge of the lobe in the top right panel, we can see features that are missing from the bottom left, \ful~, image.  These features come from higher Faraday depths than selected for this image. 
 
 %Some features from the NW side of the lobe seen on the top and bottom right panels are missing for \ful, because they come from higher depths than are being sampled here.
 Most of the same structures in \ful~ are also visible in a smoothed version of the Faraday cube, \revv{in the bottom right,} showing what would be observed at \nomi~ resolution.  The color scale is the same for \nomi~ and \ful.  However, the colors are heavily muted in \nomi~ because each tomography plane is sampling a broader range in Faraday depth.  In addition, features in the NW part of the lobe are visible in \nomi, although \ful~ shows that they are not actually present at the depths displayed. \rev{As before, there is no difference in the information available from \nomi~ and \ful, \revv{if one examines the full cubes.} only that the smaller restoring beam in \ful~ allows things to be seen more clearly.}
 
\begin{figure}
     \centering
     \includegraphics{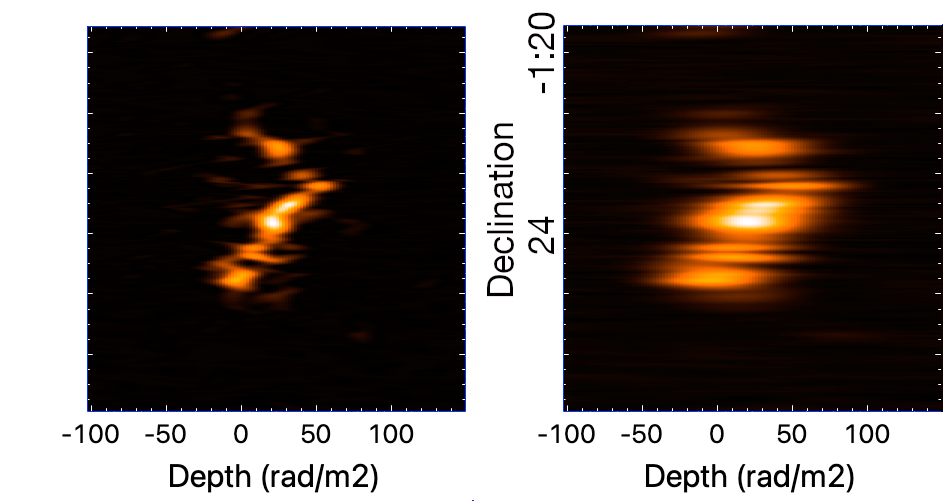}
     \caption{Polarized intensity from 3C40B's southern lobe, showing the Faraday depth distribution at each Declination, at the fixed Right Ascension of 01h25m47s. Left: \ful~ resolution. Right: smoothed to \nomi~ resolution.}
     \label{3C40Sdepth}
 \end{figure}

 The three tomography planes in the bottom left panel are a subset of those used to make the \ful~ movie of the full 3D structure of this lobe in \cite{Rudnick2022}.  The 3D structure is heavily blurred with \nomi~ resolution, as illustrated  in a single Declination vs. Faraday depth plane, \revv{at fixed R.A..} in Figure \ref{3C40Sdepth}. \rev{Note that since there is a single dominant Faraday depth at each position, the uncertainty in depth is identical for \ful~ and \nomi.  However, the changes in depth with Declination are much clearer in \ful, simply because of the smaller restoring beam.  This highlights a problem with how we display 2D information, whether here in depth vs. position space, or in a regular position-position image.  Small shifts in the centroid position may be quite significant when the signal:noise is high, but this will be obscured by using the same restoring beam as when the accuracy is lower.  This was the motivation behind the ``maximum-entropy" technique introduced by \cite{maxen}, although it is currently not being used for interferometry images.  It also motivates ``adaptive smoothing," commonly used in X-rays, as introduced by \cite{adaptX}.}

\section {Discussion}\label{discussion}

Faraday depth variations in extended sources carry information about the magnetized thermal plasmas in foreground screens, and in the medium local to the synchrotron source, including regions where the thermal and relativistic plasmas are mixed on macroscopic scales.  There are various ``figures of merit" for such studies, including the basic interferometer properties of sensitivity and angular resolution.  For the Faraday emission itself, two additional parameters of importance, the resolution in Faraday depth ($\Phi$) and the maximum detectable breadth in Faraday depth \rev{$W_{max}$}, which are set by the coverage in wavelength.\footnote{The third parameter of interest, the largest detectable Faraday depth, depends on the bandwidth of individual channels, and is often configurable in the backend receiver systems.  }

As we have shown in this paper, the \revv{commonly used} Faraday synthesis procedures do not exploit the full information available in the complex Faraday spectrum;  our goal was to explore what additional measurements are available. \revv{For both the commonly used Faraday synthesis procedure, using $\lambda_0^2=\langle\lambda\rangle^2 \approx \langle\lambda^2\rangle$ and our ``full" synthesis, using $\lambda_0^2=0$, the clean components are identical;  the essential difference between the two is the use of a narrower restoring beam for \ful, corresponding to the width of the peak in the real component of the spectrum.   To focus on the role of the restoring beams, we summarize what we've learned in terms of the two different beam widths, $\Phi_{nom}$ and  $\Phi_{full}$.  In the case of the MeerKAT L-band system, $\Phi_{nom} \approx 3\times \Phi_{full}$, with the corresponding values for other surveys summarized in Table \ref{surveys}.} 
%We thus created spectra at what we termed the ``full" resolution by setting \revv{$\lambda_0^2=0$} in the \revv{commonly used} Faraday transform, and using the width of the real component of the beam, $\Phi_{full}.$ The standard procedure,  using \revv{$\lambda_0^2=\langle\lambda\rangle^2 \approx \langle\lambda^2\rangle$}, had a 
%Faraday beam size,  $\Phi_{nom}$, which was $\sim~3\times$ larger for the MeerKAT L-band system.  The ratio of these two beam sizes depends on the detailed wavelength coverage, which varies for each survey and was summarized in Table \ref{surveys}.  For modern, wide-band systems, the ratio is typically a factor of a few.  \rev{In order to emphasize that the size of the restoring beam is the essential thing distinguishing these two procedures, we summarize what we've learned in terms of $\Phi_{full}$ and $\Phi_{nom}.$}

There are \revv{a number of} important lessons learned from these experiments. \revv{The most relevant figures for each are given in brackets.}\\
\indent $\bullet$ In the idealized case of a single Faraday depth in each pixel, both $\Phi_{full}$ 
\revv{($\approx  \frac{2}{\lambda_{max}^2 + \lambda_{min}^2}$)} 
~ and~  $\Phi_{nom}$ 
\revv{($\approx \frac{3.8}{\lambda_{max}^2 - \lambda_{min}^2} $)}
give the same results in terms of derived values and their accuracy \revv{[Figure \ref{oneRM}].}\\
\indent $\bullet$ \revv{There is a bias in the recovered amplitudes due to cleaning, which is comparable to the rms in the Faraday spectra. This needs to be simulated and corrections applied in each individual use of Faraday clean [Figure \ref{oneRM}]. }\\
\indent $\bullet$ In mapping applications, the use of $\Phi_{full}$ reduces spatial smearing in tomography images \revv{[Figures \ref{faradayexamp}, \ref{3C40Sdepth}]} and provides distinct advantages for significant regions of parameter space {\em viz., } the structure on scales between $\Phi_{full}$ and $\Phi_{nom}$, in a) tracing of spatial patterns \rev{related to Faraday depth} \revv{[Figure \ref{3C40S}]}, b) the detection of Faraday complexity, \revv{[Figures \ref{fig:resid}, \ref{residtwo}]} and c) the isolation of structures to their proper Faraday tomography image \revv{[Figure \ref{sel3395}].}\\
\indent $\bullet$ ``Spurious features", i.e., peak emission in the Faraday spectrum where no true power is present, can arise from interference between components in the complex Fourier space \revv{[Figure \ref{sum2}];} the problem is significantly worse for $\Phi_{nom}$ than it is for $\Phi_{full}$ \revv{[Figure \ref{spurious}]}. \\
\indent $\bullet$ Although Faraday complexity can be detected on scales $<\Phi_{nom}$, the detectability is a function of the phases of the underlying components,  and the details of the recovered structures are not accurate in this range \revv{[Figure \ref{twosep}].}\\
\indent $\bullet$ \revv{We introduce the quantity $W_{max}= 0.67(\lambda_{max}^{-2}+\lambda_{min}^{-2})$ \radmm, which represents the extent of a continuous distribution in Faraday depth beyond which the power in the Faraday spectrum drops by over a factor of 2 [Figure \ref{wmaxpic}]. Based on their respective values of $\frac{\lambda_{max}}{\lambda_{min}}$, most current surveys will not be able to both simultaneously resolve continuous spectra and have the sensitivity to detect them [Table 1].}
%Even their detectability is a function of the relative phases of multiple components, in addition to their extent in Faraday depth.\\
\vspace{0.1in}

Mapping applications at \rev{$\Phi_{full}$} resolution, instead of \rev{$\Phi_{nom}$}, offer perhaps the greatest potential. In the case where all variations in Faraday depth are due to patchy foreground screens, all of the useful information is found in the images of peak amplitude Faraday depth, supplemented by fractional polarization or depolarization information.  However, as has been shown by \cite{deGasperin2022} (their Fig. 16), there are coherent patterns in Faraday depth linked directly to total intensity structures in the northern relic of Abell~3667.  These imply a Faraday medium \emph{local} to the synchrotron source, and thus enables the study of the 3D structures and the relationship between the thermal and relativistic plasmas.  

Even more dramatic examples are shown by \cite{Rudnick2022}, where the 3D structure of radio filaments and lobes are shown through tomography maps and movies.  As we showed in Figures \ref{3C40S} and \ref{3C40Sdepth}, the improved {\em spatial} resolution of \rev{$\Phi_{full}$}, in the presence of Faraday variations, is critical to understanding the underlying structures.  Conversely, the standard way that Faraday results are shown in the literature is through maps of peak amplitude Faraday depth maps. {\em Even in the case of simple Faraday spectra}, with one component along each line of sight, the Faraday structures are not \rev{easily} visible (see upper left panel of Fig. \ref{3C40S} and the left panel of Fig. 16 in \cite{deGasperin2022}, \revv{unless a narrower restoring beam is used}.

Just as investigations commonly are enriched by the highest possible spatial resolution data, Faraday studies will become more powerful with better resolution in Faraday depth space.  Beyond this general argument, we can ask whether actual sources will have Faraday complexity in the newly accessible regimes.  In Figure \ref{fig:residratio}, we showed that the residual power, indicating complexity, was approximately 2$\times$ higher in \rev{$\Phi_{full}$} for continuous distributions with Faraday widths in the range 6 - 27~\radmm. In comparison, the values of $\sigma_{RM}$, derived by \cite{Osinga2022} based on the depolarization of 819 sources, fell into this range 70\% of the time, with a median value of 10~\radmm.  

In addition to the \rev{$\Phi_{full}$} reduction of spatial broadening from Faraday structure, and its ability to identify Faraday complexity, it also provides a significant improvement in the \revv{elimination} of ``spurious" features.  Such features represent the appearance of power in the Faraday spectrum at depths where no true power is present.  Instances of ``spurious" power can be seen in Figure \ref{sum2}, the simulation with two Faraday components.  Looking at the second row, e.g., we see that near zero separation (the bottom of the panel), the spectral power peaks at the middle Faraday depth, as expected.  As the separation increases, the twin peaks become dominant, again as expected.  However, as the separation increases further, there again appears power at the central depth, where none actually exists.   The separation at which this spurious power appears is a function of the relative phase of the two components.  Figure \ref{spurious} further shows that this spurious power is considerably stronger for \rev{$\Phi_{nom}$}, as opposed to \rev{$\Phi_{full}$}.  Figure \ref{228} shows spurious features where there is strong mixing within each beam;  substantial power is seen at the depth of 20~\radmm~  where none is actually present (between the black lines).  The existence of these spurious feature\revv{s} can/will confuse our interpretation of both spectra and tomography maps, and in some cases, even maps of the peak amplitude Faraday depth.

We also found that our sensitivity to continuous distributions of Faraday depth, fell off at much smaller widths than expected, \rev{and we introduce a new variable, $W_{max},$ to characterize the width at which the recovered power falls to half of the value it would have for the same input power, but a narrower width.} If the underlying emission has variations in polarization angle, or if the Faraday and synchrotron media are intermixed, then the \rev{maximum width} will be reduced \rev{still further}. With a ratio of $\frac{\lambda_{max}}{\lambda_{min}}~\sim~2$, we were unable to detect an input ``tophat"-like structure with \rev{$\Phi_{nom}$}, and only marginally with \rev{$\Phi_{full}$}.  In our survey compilation (Table \ref{surveys}), only SKA-low offers the potential to properly resolve such structures.

All of these findings suggest a cautionary approach to our interpretation of Faraday data.  Where quantitative results are necessary, it will likely be necessary to perform ``forward-modelling,"  i.e., to assume a series of underlying models, propagate them through the observing and analysis setup, and see which of  the results are consistent with the observations.  Complex observed Faraday spectra provide exceptional challenges in this regard. In source S1 in Abell~3395 shown above, some of the locations had spectra with multiple peaks, while others had simple single-peaked spectra.  Whether the multiple peaks were due to distinct components within a single spatial beam, broad Faraday distributions beyond $W_{max}$, or were spurious interference features would require extensive additional data at different wavelengths.  

There are also a number of other innovative techniques being developed for better deriving information from wideband observations.  Note that early attempts, such as fitting the Q,U spectra directly \citep{Farnsworth2011,Osullivan2012} do allow use of the full resolution, but at the expense of requiring prior knowledge of the form of the Faraday spectrum (e.g., number of Faraday thin components).  Recently, \cite{Pratley2021} have introduced a non-parametric method for Q,U fitting that circumvents this difficulty and can reconstruct more complex spectra.  \cite{Ndiritu2021} use  Gaussian Process Modeling which reduces sidelobe problems from gaps in wavelength coverage and utilizes the full resolution complex spectral information.  They demonstrated equivalent performance, e.g., to the Q,U fitting methods, but without needing the prior knowledge.  \cite{Cooray2021} use an iterative reconstruction algorithm which preserves the full resolution available.  With their simulated band spanning 300~MHz - 3000~MHz (which would require combining all three SKA1 Mid bands, e.g.), they achieve a factor of $\sim$2 in effective resolution, (see their Figure 2) as expected here using Equations \ref{phinom} and \ref{phiful}. Their reconstructions are also facilitated by the very high ratio of $\frac{\lambda_{max}}{\lambda_{min}}=10.$  

%This indicates that the Faraday screen is largely associated with the
%AGN jets.
%\begin{figure*}
%\centerline{
  %\includegraphics[height=3.5in,angle=-90.]{figs/MKBeam.ps}
  %\includegraphics[height=3.5in,angle=-90.]{figs/MKBeam_refLamb2.ps}
%  \includegraphics[width=3.5in]{figslr/dRM_depol.png}
%  }
%\caption{ This is a figure caption
%} 
%\label{dRMdepol}
%\end{figure*}
\section{Concluding remarks and future work}
Through simulations and examinations of real data, we have learned important lessons about the intrinsic reliability of Faraday spectra to recover the true underlying Faraday structure.  These lessons, summarized in Section \ref{discussion}, have important implications for our design of polarization experiments, for the interpretation of spectra and for our use of the powerful Faraday tomography techniques. 

For multiple applications, we have found that these problems are reduced, and diagnostic power is increased, by using the ``full" Faraday resolution. It is therefore important that \ful~ spectra are routinely used in the pipelines of polarization surveys, and in individual investigations.  At this stage of our knowledge, it would be prudent to produce these in parallel with \nomi~ spectra and imaging, to improve our understanding of their reliability. \rev{Since this involves changing only the restoring beam in the deconvolution process, it is trivial to implement.}

A variety of investigations should be done to extend the initial work presented here.  In particular, the processing pipeline for the POSSUM survey (van Eck et al., in preparation)  uses the quantity $\sigma_{add}$ as a measure of Faraday complexity.  $\sigma_{add}$ is the value of additional power, measured by a maximum likelihood scheme, to explain the residual fluctuations in Q$(\lambda^2)$, U$(\lambda^2)$, after subtraction of the best-fit $\delta$-function Faraday component.  It would be extremely useful to compare the detectability of complexity using the residuals in the main lobe of  \ful, as presented in Figure \ref{fig:residratio}, compared to using $\sigma_{add}$, for a variety of simulated cases.  

Further work is important to understand whether the advantages of \rev{using $\Phi_{full}$,} as shown here, also apply to other surveys, especially those with a higher ratio of $\frac{\Phi_{nom}}{\Phi_{full}}$, such as the POSSUM surveys.  It is possible that the phase instabilities which motivated \bdb~ to adopt \nomi~ will reappear when this ratio is significantly higher than the value of 2.8 as studied here. \revv{We have also done some very simple experiments to examine how the performance changes as a function of the number of Faraday depths sampled across $\Phi_{full}$.  Our tentative results are that performance is not significantly affected as long as there are at least four samples across the beamwidth.  This deserves more thorough study.}

Additional experiments probing the influence of the sidelobe magnitude on stability and the generation of spurious features, would also be of great value, and could influence how gaps in coverage are treated, whether tapering/weighting in $\lambda^2$ space is of use, etc.   All of this assumes, in addition, that spectral dependencies have been removed;  when there are multiple components within a beam, that may not be possible, and  the effects on the Faraday reconstruction  must be understood.

\rev{It would be quite useful to explore the use of variable width restoring beams for features at different S:N levels, similar to the adaptive-smoothing commonly used in X-ray imaging.}

Finally, well-designed direct comparisons of the other Faraday reconstruction techniques mentioned above, with fiducial models and realistic observing parameters (including wavelength gaps, noise variations across the band, etc.), and real data, are important and timely. Such comparisons may show that different methods are more practical/effective for different types of sources, or for surveys as opposed to individual source studies. \revv{We also note that although $\Phi_{full}$ is a ``natural" choice in some sense, not depending on some arbitrary choice of parameters, there is nothing in principle that would prevent using even smaller restoring beams.  We have not explored the associated advantages and problems here.} Given the enormous investments being made in polarization surveys, \revv{all of these types} of investigative work will provide substantial returns in scientific productivity.

\section*{Acknowledgment}
We  thank Sasha Plavin, Anna Scaife, George Heald, Cameron van Eck, Shane O'Sullivan, Miguel Carmano, Shinsuke Ideguchi, and  Yoshimitsu Miyashita for helpful comments. The anonymous referee provided critical feedback that allowed us to determine that the restoring beam was the key to the differences in results between the two methods, \revv{pointed us to the clean bias explanation for reduced recovered amplitudes, and provided a number of other useful suggestions for improving the paper.}

The MeerKAT telescope is operated by the South African Radio Astronomy Observatory, which is a facility of the National Research Foundation, an agency of the Department of Science and Innovation.  W.C.acknowledges support from the National Radio Astronomy Observatory, which is a facility of the U.S.  National Science Foundation operated under cooperative agreement by Associated Universities, Inc..
%%%%%%%%%%%%%%%%%%%%%%%%%%%%%%%%%%%%%%%%%%%%%%%%%%
\section*{Data Availability}

 The software utilized in the paper are available at http://www.cv.nrao.edu/$\sim$bcotton/Obit.html. FITS files of the various experiments will be made available upon reasonable request.  MGCLS products are publicly available (\url{https://doi.org/10.48479/7epd-w356}).



\bibliographystyle{mnras}
\bibliography{FullRes_accepted.bbl} % if your bibtex file is called example.bib

%\begin{thebibliography}{99}
%\end{thebibliography}

%%%%%%%%%%%%%%%%%%%%%%%%%%%%%%%%%%%%%%%%%%%%%%%%%%

%%%%%%%%%%%%%%%%% APPENDICES %%%%%%%%%%%%%%%%%%%%%

\appendix

\section{\rev{Depolarization from continuous Faraday distributions}}
\rev{In this section, we examine more closely the depolarizing effects of a continuous distribution of input Faraday depths, 
as illustrated in Figure \ref{fig:maxscale} for a tophat distribution.  To set the context, we recognize that for a single
Faraday component at depth $\phi_0$, the \emph{input} polarized intensity $\left|P(\lambda)\right|=\left|Q(\lambda)+iU(\lambda)\right|$ is constant (after correcting for any spectral
dependence).  All of this power is then recovered in the Faraday spectrum at $\mathcal{F}(\phi_0)$.  Similarly, for two
delta-function components at $\phi_1$ and $\phi_2$, $\left|P(\lambda)\right|$ is a sinusoid, with no monotonic decrease as a function of $\lambda$, and all of the power will be recovered by the sum of $\mathcal{F}(\phi_1)$ and $\mathcal{F}(\phi_2)$.}

The situation changes for a continuous distribution of Faraday depths, where $\left|P(\lambda)\right|$ \revv{generally} decreases monotonically with increasing $\lambda$\footnote{\revv{For example, }when there are sharp edges, as in a tophat distribution, the power will again rise past after the null, but we limit our discussion here to wavelengths below that of the first null.}, in other words, wavelength-dependent depolarization;   correspondingly, there will be reduced power in the Faraday spectrum, even after integrating over the range of $\phi$ where there is signal. 

\revv{In Eq. \ref{F_dispersion}, we wrote the Faraday spectrum in terms of $Q_j$ and $U_j$, the values of those quantities at wavelengths $\lambda_j$ (and dropping the spatial dependence).  We now decompose Q and U into a series of components each at a different Faraday depth $\phi_k$, viz.  $Q_{j}= \sum_k Q_{j,k}$ and  $U_{j}= \sum_k U_{j,k}$.  In the simplest case, if the zero wavelength polarization angle for each component, $\psi_k = 0$, and the amplitude at each Faraday depth is a constant $Q_0$, over some range $\phi_0$ to $\phi_0 + \Delta\phi$ then $Q_j = \sum_k Q_0 cos(2\phi_k\lambda_j^2)$.  We now  write this for transparency in integral form}

\begin{equation*}
 ~ ~ ~ Q(\lambda) = ~ ~  Q_0 \int_{\phi_0}^{\phi_0+\Delta\phi} cos(2\phi_k\lambda^2)\,d\phi   
\end{equation*}

%At each wavelength, there is thus a corresponding complex polarized flux $P_j = \sqrt((Q_j)^2 + (U_j)^2)e^{i\psi_j}$ .  In the simplest caseWe can also write this as and can take an initial distribution P($\lambda,\phi$) and integrate over $\phi$ to calculate the \emph{net input} power at a given wavelength, $\lambda$.  For the simplest case of a constant amplitude polarization from $\phi_0$ to \revv{$\phi_0$+$\Delta\phi$}, this can be written as}
%\begin{equation}
%\rev{\left|P_{net}(\lambda)\right| =  \left|\int_{\phi_0}^{\phi_0+\Delta\phi} P(\lambda,\phi) \,d\phi \right|}
%\label{pl}
%\end{equation}
%\rev{For simplicity, we assume a polarization angle of zero, and now look at the real component}
%\begin{equation*}
%\revv{\frac{Q_{net}(\lambda)}{Q_0} =  \int_{\phi_0}^{\phi_0+\Delta\phi} cos(2\phi\lambda^2) \,d\phi }
%\end{equation*}
\begin{equation}
\revv{~ ~ ~ ~ ~ \ \ \ \ \ \ \ \ \ \ \ =~Q_0~(sin(2[\phi_0+\Delta\phi]\lambda^2)-sin(2\phi_0\lambda^2))}
\label{ql}
\end{equation}

$Q(\lambda)$ goes to zero when $\Delta\phi=\pi$.  If we now evaluate this at $\lambda_{min}$, where the depolarization is least, we obtain
\begin{equation}
\rev{\Delta\phi = \pi/\lambda_{min}^2}
\end{equation}
\rev{which we recognize as the expression for \emph{max-scale} from \bdb.}

\revv{The same calculation, with an overall shift in phase, applies to $U(\lambda)$.} Thus, \emph{max-scale} represents the extent ~($\Delta\phi$)~ of the Faraday depth distribution at which the total power at $\lambda_{min}$, summed over all Faraday depths, $P(\lambda_{min})=\sqrt(Q(\lambda_{min})^2 + U(\lambda_{min})^2$ goes to \emph{zero}.  \revv{But for all other values of} $\lambda > \lambda_{min}$, the total power will go to zero at even smaller values of $\Delta\Phi$. Therefore, \revv{the commonly used} \emph{max-scale} is substantially larger than the value of $\Delta\phi$ where $P(\lambda)$ drops to one half when integrated over all $\lambda$. 

\revv{To estimate this latter value, we  a) summed $Q_{j,k} ~ (i.e., Q(\lambda_j,\phi_k))$ and $U_{j,k} ~  (i.e., U(\lambda_j,\phi_k))$ over an assumed distribution of $\mathcal{F}(\phi_k)$, and then b) summed $\sqrt((Q(\lambda_j)^2~ +~ U(\lambda_j)^2)$ over an assumed range in $\lambda$.}
\begin{figure}\centering
\includegraphics[width=2.5in]{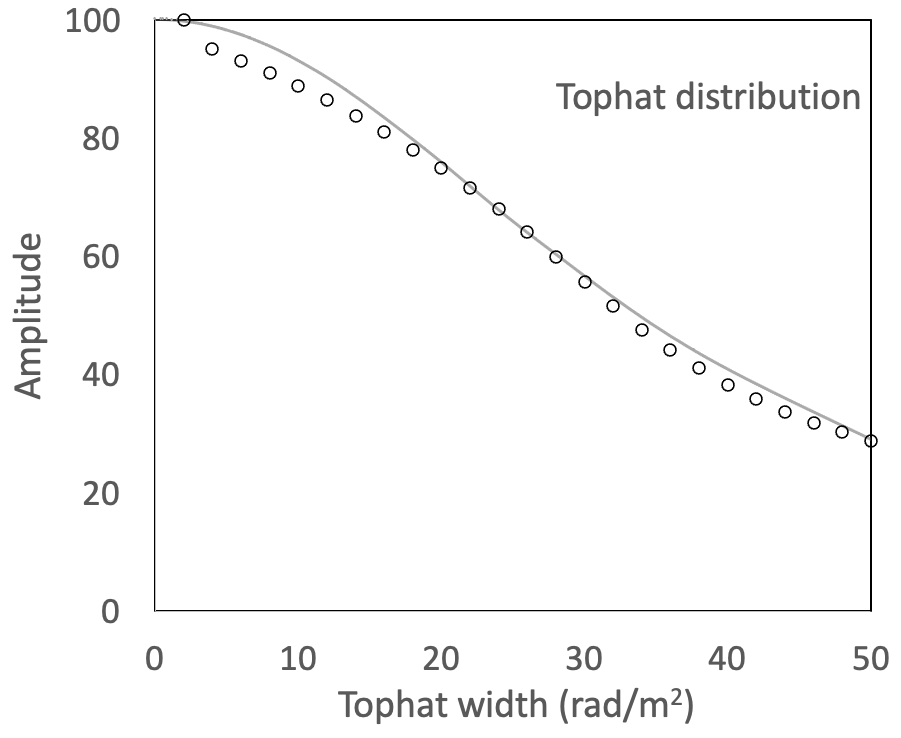}
\caption{\rev{Open circles indicate the power summed over the full Faraday spectrum for tophat distributions of different widths.  The solid curve is the calculated $P_{net}$, \emph{input} power as a function of the tophat distribution width. Both are normalized to 100 at zero width.}}
\label{tophcalc}
\end{figure}
\rev{The results of this calculation, as a function of the assumed $\Delta\phi$ for a tophat distribution, are shown in Figure \ref{tophcalc}. The data points  represent the summed power in the \ful~  Faraday spectrum, assuming the MeerKAT L-band wavelength coverage with no gaps.  It is thus similar, but not identical, to the tophat results with coverage gaps  in Figure \ref{fig:maxscale}.  \emph{We thus see that the recovered power in the full Faraday spectrum is well-matched to the input power.}  }

\rev{In practice, however, the recovered power in the spectrum will be less than shown here.  While we summed the power of the entire noise-free Faraday spectrum,  in the real case, one would need to choose a region over which to integrate, and power that had been scattered into distant sidelobes would be lost.  In addition, as shown in Figure 7, the Faraday power shows up in the wings of the input distribution, where they may be difficult to recognize.  Finally, any variations in polarization angle at the different Faraday depths will further reduce the observed power in the spectrum.  \emph{Thus, our calculation should be considered a strict upper limit to the recoverable power.}}

\rev{It is often assumed that the distribution of Faraday depths will follow a Gaussian distribution, so we have repeated the above calculation as a function of the FWHM of the Faraday depth distribution.   We did this for a wide variety of assumed wavelength coverages, in order to determine the value of the FWHM at which the input power (and thus the maximum recoverable power in the Faraday spectrum) fell by a factor of 2.   We call this value $W_{max}$, and recommend it, as opposed to \emph{max-scale}, as an indicator of the sensitivity to continuous Faraday distributions.  We have listed $W_{max}$ for each of the surveys listed in Table \ref{surveys}.   }

\begin{figure*}
\includegraphics[width=3.4in]{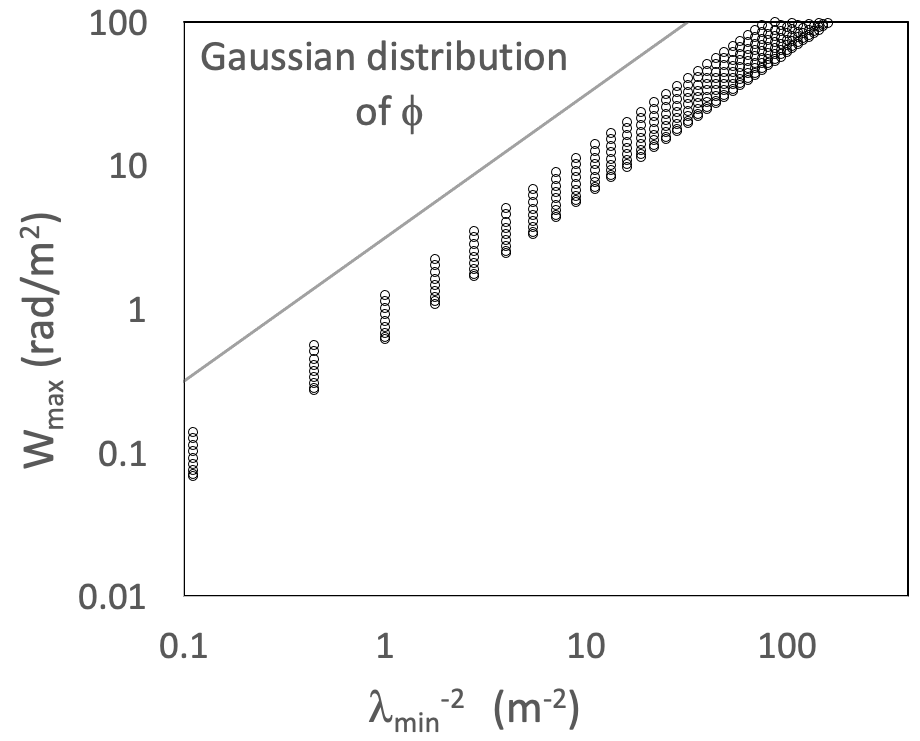}
\includegraphics[width=3.4in]{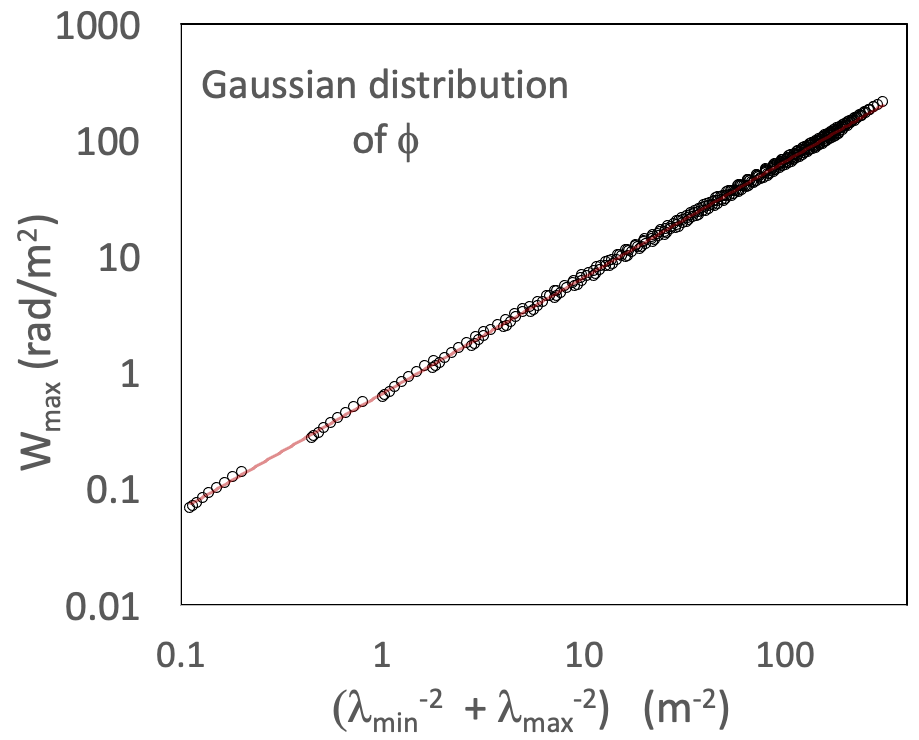}
\caption{\rev{$W_{max}$, the FWHM of the Gaussian distribution at which $P_{net}$ falls by a factor of 2,  as a function of wavelength coverage.  Left: plotted as a function of $\lambda_{min}^{-2}$ only, ignoring $\lambda_{max}^{-2}$. Solid line shows \emph{max-scale}. Right: Including dependence on both $\lambda_{min}$ and $\lambda_{max}$, showing the relationship fit by $W_{max}= 0.67(\lambda_{min}^{-2}+\lambda_{max}^{-2}) ~ rad~m^{-2}.$}}
\label{wmaxpic}
\end{figure*}

\rev{In order to determine how $W_{max}$ depends on the wavelength coverage, we plot it in two different ways in Figure \ref{wmaxpic}.   First, we plot $W_{max}$ vs. $\lambda_{min}^{-2}$, which is the scaling  of \emph{max-scale}.  We find the expected overall behavior, but see that different trends are found depending on the value of $\lambda_{max}^{2}$.  The second plot shows $W_{max}$ vs. the quantity ($\lambda_{min}^{-2} + \lambda_{max}^{-2})$; we find that all the results are consistent with the single trend}
\begin{equation}\label{wmax}
%\sigma_{0.5}=0.29 (\lambda_{min}^{-2} + \lambda_{max}^{-2}) = 0.29\lambda_{min}^{-2}(1~+~\rho^{-2}),
\rev{W_{max} = 0.67 (\lambda_{min}^{-2} + \lambda_{max}^{-2}) = 0.67\lambda_{min}^{-2}(1~+~\rho^{-2}) rad~m^{-2},}
\end{equation}
\rev{where $\rho = \frac{\lambda_{max}}{\lambda_{min}}$.  $W_{max}$ is the proper quantity, for a given wavelength coverage, to characterize when the power will drop by a factor of 2.  It is important to note that these values are considerably less than \emph{max-scale}, which will affect modeling of physical systems and the planning of observations. }
%}
The equivalent relationship for a tophat distribution of width W is
\begin{equation}
W_{max}=0.45 (\lambda_{min}^{-2} + \lambda_{max}^{-2}) = 0.45\lambda_{min}^{-2}(1~+~\rho^{-2}) ~ rad~m^{-2}
\end{equation}
which is  $\approx30\%$ different than the Gaussian case.  Since the actual input and recovered powers also depend on other factors such as the polarization angle variations across the Faraday depth distribution, we recommend that Eq. \ref{wmax} be used as a reasonable approximation for the FWHM of the distribution at which the power drops by a factor of 2. 
%}
\end{document}